\documentstyle[aps,epsf,floats,twocolumn,prb,euscript,eufrak]{revtex}

\def\etal{~\textit{et~al.}} 
\def\={\!=\!}    
\def\ra{\rangle} 
\def\la{\langle} 
\def\const{{\mathrm const}}       
\def\sgn{{\mathrm sgn}}           
\def\num{{\textit{(num.fact.)}}}  
\newcommand{\mb}[1]{{\mathbf #1}} 
\newcommand{\bfgr}[1]{{\boldmath{\mbox{$#1$}}}}

\def\br{\mb r}         
\def\hbx{\hat\mb x}    
\def\hby{\hat\mb y}    

\def\mt{{\hat t}}      
\def\FA{{\mathfrak A}}          
\def\FB{{\mathfrak B}}          
\def\FC{{\mathfrak C}}          
\def\FD{{\mathfrak D}}          
\def\cvr{{\EuScript M}}         
\def\benrg{{\epsilon}}          
\def\EE{{\EuScript E}}          
\def\EEgs{\EE^{(gs)}}           
\def\Eoptdef{{\Large\mbox{$\Huge\mathfrak e$}}_{\mathrm opt.def.}} 
\def\FZ{{\mathfrak Z}}          
\def\conduct{\overline\sigma}                     
\def\rlyap{{\EuScript\, P}_{\mathrm Lyap}}        
\def\rlyapz{{\EuScript\, P}_{\mathrm Lyap, 0}}    
\def\lnL{\,\ln L\,}                
\def\Gopt{{\Gamma_{\mathrm opt}}}  
\def\Prob{{\textit Prob\,}}        
\def\CA{{\mathcal A}} 

\begin{document}
\twocolumn[\hsize\textwidth\columnwidth\hsize\csname@twocolumnfalse%
\endcsname
\draft

\title{Particle-hole symmetric localization in two dimensions}
\author{Olexei~Motrunich,$^1$ Kedar~Damle,$^2$ and 
David~A.~Huse$^1$}
\address{
$^1$ Physics Department, Princeton University, Princeton, NJ 08544\\
$^2$ Physics Department, Harvard University, Cambridge, MA 02138
}
\date{\today}
\maketitle

\begin{abstract}
{We revisit two-dimensional particle-hole symmetric sublattice 
localization problem, focusing on the origin of the observed 
singularities in the density of states $\rho(E)$ at the band center 
$E\!=\!0$.  The most general such system 
[R.~Gade, Nucl. Phys. B {\bf 398}, 499 (1993)] exhibits critical 
behavior and has $\rho(E)$ that diverges stronger than any 
integrable power-law, while the special 
{\it random vector potential model} of 
Ludwig\etal~[Phys. Rev. B {\bf 50}, 7526 (1994)]
has instead a power-law density of states with 
a continuously varying dynamical exponent.  We show that the latter 
model undergoes a dynamical transition with increasing 
disorder---this transition is a counterpart of the static 
transition known to occur in this system; in the strong-disorder 
regime, we identify the low-energy states of this model with the
local extrema of the defining two-dimensional Gaussian random 
surface. 
Furthermore, combining this ``surface fluctuation'' mechanism 
with a renormalization group treatment of a related vortex glass 
problem leads us to argue that the asymptotic low $E$ behavior of 
the density of states in the {\it general} case is
$\rho(E) \sim E^{-1} e^{-|\ln E|^{2/3}}$, 
different from earlier prediction of Gade.
We also study the localized phases of such particle-hole symmetric 
systems and identify a Griffiths ``string'' mechanism that
generates singular power-law contributions to the low-energy 
density of states in this case.
}
\end{abstract}
\vskip 0.3 truein
]

\section{Introduction}
\label{sec:INTRO}
This article is devoted to a careful study of the localization
properties of certain two dimensional systems with some
special symmetry properties that make them stand apart from
the other more generic universality classes of Anderson
localization in two dimensions.
The models we consider are simply tight-binding models 
of noninteracting particles moving on {\it bipartite} lattices
in two dimensions (2D).\cite{Gade}  
The hopping matrix elements are random, but there are no on-site 
potentials, so the spectrum is strictly particle-hole symmetric.  
Such 2D localization problems have received particular 
attention\cite{GadWeg,HikShiWeg,LudMpafShaGri}
in the context of the integer quantum Hall plateau transition 
studies, and exhibit critical delocalized states only at the special 
energy $E \!=\! 0$.  At all nonzero energies the eigenstates are
exponentially localized with a finite localization length.

Unlike the quantum Hall systems, which have an analytic density
of states (DOS) at the corresponding delocalized critical points
in the spectrum, the bipartite random hopping (BPRH) models have 
a strongly singular DOS at the band center.  
This unusual feature makes the sublattice problems very interesting 
in their own right since the disorder evidently has a very dramatic 
effect in the presence of the particle-hole symmetry.

Recent studies of quasiparticle localization in dirty 
superconductors, where a somewhat analogous particle-hole 
symmetry appears naturally in the Bogoliubov-de Gennes formulation,
have resulted in renewed interest in such special localization 
problems.  For instance, the case with broken spin-rotation 
invariance\cite{SenMpaf:noSR} can be formulated as a
pure imaginary random hopping (ImRH) problem, and such ImRH
problems constitute the simplest particle-hole symmetric 
generalization of the real BPRH models, obtained from 
the corresponding real BPRH by allowing additional 
{\it purely imaginary} hopping amplitudes between the sites of 
the same sublattice and performing a gauge transformation on 
one sublattice to make the inter-sublattice couplings imaginary.
In the generic case, such systems all show some enhancement in
the quasiparticle density of states at the band center, 
while the special case of a spinless superconductor with 
time reversal invariance (which constrains the couplings within 
a sublattice to be zero) 
maps precisely onto a real bipartite localization problem with
all the concomitant strong spectral singularities.

To put our analysis of the bipartite models in context,
it is useful to begin with a quick summary of what is known 
in the literature:
Ludwig\etal\cite{LudMpafShaGri} showed that a special case
of the sublattice problem, the {\it random vector potential model} 
(see Sec.~\ref{sec:DEFS:longtd} below),
has a power-law density of low (i.e., near zero) energy states,
\begin{equation}
\rho(E) \sim E^{-1+2/z}~,
\label{rho:griff}
\end{equation}
parameterized by a continuously varying dynamical exponent $z$.
If we define an energy-dependent length scale $L(E)$ from 
the density of states so that the integrated density 
$N(E) \equiv \int_0^E dE' \rho(E') \equiv L^{-2}(E)$,
then Eq.~(\ref{rho:griff}) corresponds to dynamical scaling 
of the form $E \sim L^{-z}$.

In related work, Gade and Wegner\cite{Gade,GadWeg} developed a 
field theoretic description of the {\em general sublattice problem}
(with {\em random mass} terms in addition to random vector potential disorder---see Sec.~\ref{sec:DEFS:longtd} below), 
predicting a strongly divergent density of states
\begin{equation}
\rho(E) \sim \frac{\,1\,}{\,E\,}\, e^{-c |\ln E|^{1/x}},
\label{rho:gade:x}
\end{equation}
with $x=2$, stronger than any finite-$z$ power law 
Eq.~(\ref{rho:griff}) [these results have also been rederived
very recently\cite{GurLecLud} from a somewhat different field 
theoretic analysis].  This ``Gade form'' corresponds to a kind of 
{\em activated} (infinite-$z$) dynamical scaling
\begin{equation}
|\ln E| \sim (\ln L)^x ~,
\label{length:gade:x}
\end{equation}
when $x \!>\! 1$.

Finally, it is useful to note that these bipartite problems are 
closely related to a well-studied vortex 
glass problem.\cite{CarOst,TonDiv,HwaDsf}

The field-theoretic analyses that lead to these predictions, 
however, do not provide a direct physical picture of the origin of 
the divergent DOS at the band center and the nature of the low 
energy states.  Developing such a picture is the main thrust of the 
present work.  In particular, our analysis leads us to suggest that 
the asymptotic scaling between the energy and length in the general 
sublattice problem has the infinite-$z$ form 
Eqs.~(\ref{rho:gade:x})~and~(\ref{length:gade:x}),
but with a different exponent
\begin{equation}
x=\frac{3}{2}~.
\label{newx}
\end{equation}

In the remainder of this section, we now outline our physical 
picture of the low-energy states, and summarize the basic argument 
that leads us to Eq.~(\ref{newx}); for details and some relevant 
definitions, see Secs.~\ref{sec:DEFS}~and~\ref{sec:RG} below.

The mechanism by which disorder produces a pile-up of low-energy 
states is best understood by first considering the special 
random vector potential model.\cite{LudMpafShaGri}  
In the lattice version of this model, the hopping matrix elements 
are not independent random variables.  Instead, they are given by 
the differences between the values of a random Gaussian field 
$\Phi(\br)$ [see defining Eqs.~(\ref{tPhi})~and~(\ref{Phi})] on the 
two sites. 
This field has the statistics of a rough surface, with the variance 
of the difference in $\Phi$ between adjacent sites being 
set by a disorder strength parameter $g$.  
As we argue later in Sec.~\ref{sec:RG}, the low energy states
at strong randomness \mbox{$g \gg 1$} ``live'' near the extrema 
of the surface $\Phi(\br)$, and the logarithms of their energies 
are roughly given by the corresponding relative surface heights.  
Such extremal properties of two-dimensional Gaussian surfaces 
are well-characterized (see Appendix~\ref{app:2dGAUSS} 
for a summary of the relevant results); 
in particular, in a sample of size $L$, the corresponding
prediction for the energy of the lowest state is
\begin{equation}
|\ln E| \sim \Big[ \Phi_{\max}(L) - \Phi_{\min}(L) \Big] 
\sim \sqrt{g} \ln L ~,
\label{intro:long:start}
\end{equation}
implying power-law dynamical scaling with the dynamical exponent 
$z$ growing as
\begin{equation}
z \sim \sqrt{g}
\label{intro:zg:strong}
\end{equation}
for strong randomness $g \!\gg\! 1$.
As mentioned earlier, Ref.~\onlinecite{LudMpafShaGri} 
indeed predicts a power-law DOS in this model, but with 
the dynamical exponent
\begin{equation}
z=1+\frac{g}{\pi}~;
\label{intro:zg:weak}
\end{equation}
within the field theoretic analysis, this result appears to be 
perturbatively exact to all orders in $g$.  However, the above 
``surface argument'' shows that Eq.~(\ref{intro:zg:weak}) 
cannot hold for strong disorder, 
since the smallest energy in the problem cannot fall below 
the limits fixed by the extrema of the surface.  
Indeed, we argue in Sec.~\ref{sec:BOUNDS:longtd} that there is 
actually a dynamical transition at $g=g_c\equiv 2\pi$; 
at this transition, the exponent $z$ changes its behavior from 
the weak-disorder form~(\ref{intro:zg:weak}) to the strong-disorder 
form~(\ref{intro:zg:strong}) [see also Eq.~(\ref{zg:strong})].  
This dynamical transition is a counterpart of a static 
transition\cite{CasChaFraGolMud,CarDou01} known to occur
when the wave function $e^{-[\Phi(\br)-\Phi_{\min}]}$ 
(i.e., the zero-energy ``pseudo-eigenstate'' defined so that its peak
value is $1$) becomes normalizable in the limit $L \to \infty$.

Turning to the general case where the hopping matrix elements 
are independently random, we thus expect some power-law density 
of states to be produced as long as there is some 
``vector potential'' component to the randomness, as there is.
Moreover, a field theoretic renormalization group (RG) 
analysis\cite{GurLecLud,CarOst,TonDiv,HwaDsf} shows that any 
amount of a more general ``random mass'' disorder 
(see Sec.~\ref{sec:DEFS:continuum} below) generates additional 
vector potential randomness, driving the latter to stronger 
disorder at larger length scales and lower energy scales:
\begin{equation}
g_{\mathrm eff}(L) \sim \ln L ~.
\label{intro:gen:start}
\end{equation}

The nature of the low-energy spectrum and the corresponding 
dynamical scaling in the general case can now be heuristically 
understood by considering an ``effective'' random vector potential 
model with a scale dependent disorder strength parameter 
$g_{\mathrm eff}(L)$ given by Eq.~(\ref{intro:gen:start})
[see Sec.~\ref{sec:BOUNDS:general} for a more precise argument 
from a different perspective using the results of 
Ref.~\onlinecite{ZenLeaDsf} for a related vortex glass problem].
The low-energy states in this effective problem are again associated 
with the extrema of some effective surface (which may be identified 
with the logarithm of the magnitude of the
lowest-energy wave function), but the height extrema of this 
surface, and correspondingly the log-energies, now scale with $L$ as
\begin{equation}
|\ln E| \sim z_{\mathrm eff}(L) \ln L \sim 
\sqrt{g_{\mathrm eff}(L)}\,\ln L \sim (\ln L)^{3/2} ~,
\label{intro:gen:stop}
\end{equation}
which is the proposed ``modified-Gade form''.  
[Note that the original Gade form would obtain if the weak-disorder
form for $z$, Eq.~(\ref{intro:zg:weak}), were used:
\begin{equation}
|\ln E| \sim z_{\mathrm eff}(L) \ln L \sim^{\!\!\!\!\!\! \mbox{!?!}} 
g_{\mathrm eff}(L) \ln L \sim (\ln L)^2~,
\end{equation}
However, since $g_{\mathrm eff}(L)$ grows indefinitely
with $L$, this form cannot be used to obtain the true asymptotic 
behavior.]

So far, we have only alluded to {\it critical} bipartite systems 
whose zero-energy states are not exponentially localized.  
However, it is possible to drive such a system into a different 
{\it localized} phase keeping the particle-hole symmetry intact:  
This can be achieved, for example, by making some prescribed bonds 
that produce a complete dimer cover of the lattice stronger on 
average, thus introducing a preferred ``dimerization'' pattern.  
If this dimerization is weak, the system may remain 
critical, but strong dimerization will eventually drive it
into a localized (insulating) state.  We have found an interesting 
mechanism whereby disorder generates power-law contributions to the 
low-energy DOS in such bipartite {\it insulating} phases:  
The low-energy states are associated with the end-points of 
``strings'' along which the background dimer pattern is broken as 
in~Fig.~\ref{string}.  
Estimating the relevant probabilities shows that this mechanism is 
actually operative for BPRH models in any dimension, as well as 
in the localized phases of the related ImRH problems 
(in the latter case, particular dimerization patterns appear quite 
naturally when describing disordered superconductors with broken 
spin-rotation invariance~\cite{MotrDamHus:noSR}). 
In our bipartite systems, this ``Griffiths-like'' mechanism 
can produce a power-law divergent DOS and dominate over all 
other mechanisms of filling the band gap near zero energy 
in such localized phases.

Having introduced our principal results, we now conclude this
section with an outline of the rest of the paper:  
Sec.~\ref{sec:DEFS} defines our models and reviews the connection 
with the vortex-glass problem. 
Sec.~\ref{sec:RG} introduces a strong-randomness picture of 
the low-energy physics, motivating the more precise bounds
of Sec.~\ref{sec:BOUNDS} on the dynamical scaling in the system.
Sec.~\ref{sec:NUMERICS} summarizes our numerical evidence 
in support of our analytical arguments.  
Finally, Sec.~\ref{sec:discus} touches upon some unresolved 
questions and prospects for future study.

\section{Models, definitions, and the connection with dimers}
\label{sec:DEFS}
A bipartite random hopping problem is completely specified by
a single-particle Hamiltonian
\begin{equation}
\hat H = \sum_{\la\alpha \beta\ra} 
\Big( t_{\alpha\beta} |\alpha\ra \la\beta| + {\mathrm H.c.} 
\Big) \equiv 
\pmatrix{ 0 & \mt_{AB} \cr
          \mt^\dagger_{AB} & 0}~,
\label{H}
\end{equation}
where $\alpha$ and $\beta$ belong to sublattices $A$ and $B$ 
respectively of some bipartite lattice.  In this article, 
we only consider real  hopping problems in two dimensions, 
focusing attention on the following specific models: 
{\it $\pi$-flux}, {\it honeycomb}, and {\it general} bipartite 
lattice models.  

Each of these models is defined as a nearest-neighbor hopping 
problem on an appropriate lattice, with some additional constraints 
on the allowed signs of the couplings:
The $\pi$-flux model\cite{HatWenKoh} is defined on a square 
lattice with a requirement that there is one half of a magnetic
flux quantum per square plaquette.  A convenient gauge is 
\begin{equation}
t_{\,\mb j,\, \mb j + \hbx} = (-1)^{j_y} t_x(\,\mb j)~,
\qquad
t_{\,\mb j,\, \mb j + \hby} = t_y(\,\mb j) ~,
\end{equation}
where $\mb j \!=\! \{j_x, j_y\}$ labels sites of the square
lattice, $\hbx$ and $\hby$ are unit lattice vectors,
and $t_x(\,\mb j)$ and $t_y(\,\mb j)$ are nonnegative hopping 
amplitudes.  
The honeycomb (brickwall) lattice model\cite{fdmh88} 
is defined on a honeycomb lattice with nonnegative hopping 
amplitudes.
Finally, by the general lattice model, we mean a 
nearest-neighbor hopping model on a square or honeycomb lattice 
with no constraint on the allowed signs.

Naturally, the pure-system spectra in the above models can be quite 
different:  The pure $\pi$-flux and honeycomb lattice models have 
Dirac points and a linearly vanishing density of states at $E=0$, 
while the pure rectangular lattice model has a Fermi surface with
a smooth DOS near $E=0$ in the general anisotropic case 
and a van~Hove singularity in the isotropic square lattice case.
However, all these models are believed to have similar 
low-energy long-wavelength localization physics in the presence of
randomness.   In our studies, it is often convenient to work with 
a particular model when using a particular approach, and we switch
among the models a fair amount in what follows.

\subsubsection{Continuum description of models}
\label{sec:DEFS:continuum}
The pure $\pi$-flux model has two Dirac points.  The corresponding 
weakly-disordered model is described by a continuum Dirac 
Hamiltonian, which may be conveniently written as\cite{HatWenKoh}
$\hat h_4 = \pmatrix{ 0 & \hat h_2 \cr
                     \hat h_2^\dagger & 0}$, with
\begin{eqnarray*}
\hat h_2  \!=\! \sigma_x (-i v_x \partial_x + i A_x) +
                \sigma_y (-i v_y \partial_y + i A_y) -
                i V + M \sigma_z.
\end{eqnarray*}
In the above, the slowly varying vector and scalar potentials are
implicitly defined by
$2 t^{(0)}_x \!=\! v_x$,
$2 t^{(0)}_y \!=\! v_y$, 
$-2 \delta t_x(\,\mb j) \!=\! 
    A_x(\,\mb j)(-1)^{j_x+j_y} +V(\,\mb j)(-1)^{j_x} + \dots$,
$-2 \delta t_y(\,\mb j) \!=\! 
    A_y(\,\mb j)(-1)^{j_x+j_y} +M(\,\mb j)(-1)^{j_y} + \dots$,
and the lattice constant has been set to unity. 
A similar continuum Hamiltonian obtains for the honeycomb lattice 
model. 

The Dirac Hamiltonian $\hat h_2$ is non-hermitian and contains 
a real random mass $M$, a random imaginary potential $-iV$, 
and a random imaginary gauge field.  The $\{M,V\}$ terms
form a ``complex random mass'' part, which we refer to simply
as the random mass part of the disorder.  
In its absence, $M \!\equiv\! V \!\equiv\! 0$, 
the full Hermitian Hamiltonian $\hat h_4$ decomposes into two 
blocks, with the ``1-4'' block having a form 
\begin{equation}
\hat h_\CA = \sigma_x (-i \partial_x + \CA_1) +
             \sigma_y (-i \partial_y + \CA_2) ~,
\label{hA}
\end{equation}
identical to the (real) random vector potential model of 
Ref.~\onlinecite{LudMpafShaGri} 
[here $\{\CA_1,\CA_2\} \!\equiv\! \{A_y,-A_x\}$], 
and similarly for the ``2-3'' block.  
Here and henceforth, we will therefore refer to the
$\{A_x,A_y\}$ part of $\hat h_4$ 
as the random vector potential part of the disorder.

\subsubsection{The random vector potential model on a lattice}
\label{sec:DEFS:longtd}
Consider a ``random-surface'' BPRH model constructed from
a field $\Phi(\br)$ and a pure system Hamiltonian with bare 
hopping amplitudes $\mt^{\,(0)}_{AB}$ as follows:
\begin{equation}
t_{\alpha\beta} = e^{\Phi(\alpha)} \,\, t^{(0)}_{\alpha\beta} \,\,
                  e^{-\Phi(\beta)}.
\label{tPhi}
\end{equation}

One can easily see that if $\vec \Psi^{(0)}_A$ is a zero-energy 
eigenstate of the pure model, then $\vec \Psi_A$ 
with $\Psi_A(\alpha) \!=\! e^{-\Phi(\alpha)} \Psi^{(0)}_A (\alpha)$ 
is a zero-energy eigenstate of the random-surface model.
Moreover, it is easy to see that this construction provides a 
particular lattice realization of the random vector potential model,
if the pure system one starts with is the non-random $\pi$-flux 
model on a square lattice.  Indeed, in the continuum limit, 
we obtain
$M \!\equiv\! V \!\equiv\! 0$, 
$A_x \!=\! v_x \partial_x \Phi(\br)$, and 
$A_y \!=\! v_y \partial_y \Phi(\br)$, so that $\hat h_\CA$ of
Eq.~(\ref{hA}) becomes
$\hat h_\CA = \sigma_x (-i \partial_x + \partial_y \Phi) +
              \sigma_y (-i \partial_y - \partial_x \Phi)$.
Our random surface $\Phi(\br)$ thus represents the physical 
(non-gauge) degrees of freedom of the random vector potential, 
and we choose it to be Gaussian
\begin{equation}
{\mathrm Prob} \big[\Phi\big] \,\propto\, 
\exp\left[-\frac{1}{2g} \int\! d^2\br \, (\bfgr{\nabla} \Phi)^2 
          \right]~,
\label{Phi}
\end{equation}
with dimensionless disorder strength $g$; the relevant properties 
of such surfaces in two dimensions are summarized in 
Appendix~\ref{app:2dGAUSS}.

A remark is in order here:  In the continuum limit of the general 
problem, we have seen that there is a clean separation of the
disorder into its random vector potential and random mass parts.  
However, in the original model on a lattice, such a separation 
is not at all obvious---the only precise statement we make 
at the lattice level is that a model has purely vector potential 
randomness when it can be put in the form given by Eq.~(\ref{tPhi}).
Renormalizing on a lattice in Sec.~\ref{sec:RG}, we will still speak 
of the flows of the two parts of the disorder, implicitly assuming 
some coarse-grained description and relying on the corresponding 
results for the continuum model.  A more precise lattice alternative
to such a hybrid RG approach is the domain wall energetics 
picture of Ref.~\onlinecite{ZenLeaDsf}, which we will have occasion 
to appeal to in Sec.~\ref{sec:BOUNDS:general}.

\subsubsection{Connection with random dimer problems}
\label{sec:DEFS:dimer}
There is a direct connection between the $\pi$-flux random hopping 
problem and a random dimer model on the same square lattice.  
The dimer model consists of all complete coverings (matchings)
$\{\cvr\}$ of the (bipartite) lattice by the nearest-neighbor 
dimers; the energy of a given covering $\cvr$ is
\begin{equation}
\EE_d[\cvr] \equiv 
\sum_{\la \alpha\beta \ra \in \cvr} \benrg_{\alpha\beta},
\label{Edimer}
\end{equation}
where the sum is over all dimers in this covering, and 
$\benrg_{\alpha\beta}$'s are the random bond energies.
The fermion-dimer connection is stated most easily for
open boundary conditions:  In this case, the partition function 
of the dimer model 
$Z_d \equiv \sum_{\cvr} \exp\big( \!-\!\EE_d[\cvr]/T_d \big)$ 
at a given dimer temperature $T_d$ can be written as a determinant 
of an appropriate connection matrix $\mt_{AB}$, 
$Z_d[A, B] = \det\mt_{AB}$, where
\begin{equation}
t_{\alpha\beta} = \pm e^{-\benrg_{\alpha\beta}/T_d}~,
\end{equation}
and the signs of the hopping amplitudes are chosen exactly
as in the $\pi$-flux model.~\cite{Kasteleyn,Montroll}
The honeycomb lattice model with nonnegative bonds has a similar
dimer connection (our definitions of the $\pi$-flux and honeycomb 
models were made precisely with this in mind), but the more general 
lattice model has no such direct connection.

\subsubsection{Localized bipartite hopping systems}
\label{sec:DEFS:localzd}
The commonly studied bipartite hopping models (e.g., all of 
the above models with some generic distributions of random hopping 
amplitudes) are found to be {\it critical}, by which we mean that
the behavior at $E=0$ is neither truly delocalized nor truly 
(exponentially) localized.  Abusing language somewhat, 
we will often refer to a system in such a critical state as being 
in a ``delocalized'' or ``metallic'' phase---this serves to 
emphasize the distinction between such a phase and truly 
localized insulating phases that obtain, e.g., by introducing 
strong dimerization in the couplings.  The pure-system counterparts 
of these localized phases are typically gapped band insulators.  
In our studies, we will specifically consider 
only one such localized phase produced by introducing ``staggered'' 
dimerization on the square lattice (see Fig.~\ref{string});
the honeycomb lattice version of this is 
shown in Fig.~\ref{brickwall}(a).

\section{On the origin of low-energy states}
\label{sec:RG}
We begin by noting that the Gade-like dynamical scaling 
Eq.~(\ref{length:gade:x}) with $x \!>\! 1$ suggests that 
the effective value of disorder scales to infinity at 
asymptotically low energies (dynamical exponent $z \!=\! \infty$), 
albeit very weakly.  This motivates us to try a strong-randomness 
RG description.\cite{MotrDamHus:dyn:long,MotrDamHus:noSR}

\subsection{Strong-randomness RG}
\label{sec:RG:rg}
Consider the bipartite hopping Hamiltonian~(\ref{H}).  
The eigenstates of $\hat H$ occur in pairs with energies 
$\pm E$, and the strong-randomness RG proceeds by eliminating, 
at each step, such a pair of states with energies near the top 
and the bottom of the band:  One finds the largest (in absolute 
value) coupling in the system, say $t_{1,\,2}$ connecting sites 
$1\!\in\! A$ and $2\!\in\! B$; this defines the bandwidth 
$2\Omega = 2\max\{|t_{\alpha,\,\beta}|\}$.  
If the distribution of the couplings is broad, the two eigenstates 
of the two-site problem $\hat H[1,2]$ with energies $\pm\Omega$ 
will be good approximations to eigenstates of the full $\hat H$, 
since the couplings $t_{1,\,\beta}$ and $t_{2,\,\alpha}$ of 
the pair to the rest of the system will typically be much 
smaller.  These couplings can then be treated perturbatively, 
and eliminating the two high-energy states living on the sites 
$1$ and $2$ gives us the following effective couplings between 
the remaining sites:
\begin{equation}
t'_{\alpha,\,\beta} = t_{\alpha,\,\beta} - 
t_{\alpha,\, 2} \,\, t_{1,\,2}^{-1} \,\, t_{1,\,\beta}~.
\label{RGrule}
\end{equation}
The renormalized Hamiltonian is again a bipartite hopping problem, 
but with two fewer sites; in particular, no diagonal terms are 
generated.  This procedure is then iterated to reach lower
and lower energy.

As a first attempt, we implemented this RG numerically in both
localized and metallic phases.  
In the localized phases, the RG indeed provides a qualitatively
and quantitatively accurate description of the low-energy physics, 
with the relative ``decimation errors'' diminishing quickly at 
low energies; in this case, the RG is thus a consistent scheme 
that correctly describes the Griffiths effects analyzed in detail 
below.  
However, in the delocalized phase, the observed flow to stronger 
disorder is very weak, and the consistency of the RG is less 
clear.  We do not address in detail here this question of 
self-consistency (that is, in which regimes---if at all---the 
system flows to stronger disorder, with the RG becoming more and 
more accurate).  Rather, in the metallic phase, we view the RG
as providing a heuristic picture that complements the more precise 
arguments we present later.
Note, however, that the RG actually contains some exact and useful 
information for any strength of the disorder:~\cite{MotrDamHus:noSR}
The rule~Eq.~(\ref{RGrule}) is an exact transformation 
for the zero-energy wave function, and as such provides useful 
information on the $E \= 0$ localization properties.
Also, running this RG corresponds to employing the Sturm sequence 
method for calculating the integrated density of states at 
$E \= 0$ using a particular order that is, without any 
a priori knowledge, numerically most reliable; the intermediate 
terms that appear in this process at a particular length scale 
give us a rough idea about the corresponding energy scale 
in the problem.

\subsection{On the origin of states}
\label{sec:RG:orig}
The strong-randomness RG associates the low-energy states 
of the original Hamiltonian with the sites that ``survive'' 
down to the corresponding energy.  Such a ``free'' 
site---say, an $A$-site---is found if all the $B$-sites 
in its immediate neighborhood happen to be ``locked'' 
(by stronger bonds) into pairs with some other $A$-sites.  
Then, this free $A$-site can only couple to the next available 
$B$-sites, which are far away, and the corresponding effective 
couplings are relatively weak since they are mediated 
by a substantial intervening pair-locked region.  

It is at first difficult to see how such free sites and
the corresponding low-energy states can be produced often enough in 
dimensions $d>1$:  Naively, the energy scale associated with a pair
$\{\alpha,\,\beta\}$ of ``isolated'' sites distance $l$ apart 
is $\sim\! e^{-c_E l}$.  Now, it may seem that to produce such 
a pair we need of order $l^d$ specific events---for example, 
we can imagine having a particular pairing pattern for all the 
other sites in the neighborhood of the two unpaired sites 
$\alpha$ and $\beta$---and this has a very low probability 
$\Prob \sim e^{- c_p l^d}$ in $d>1$.  

\begin{figure}
\epsfxsize=2.6in
\centerline{\epsffile{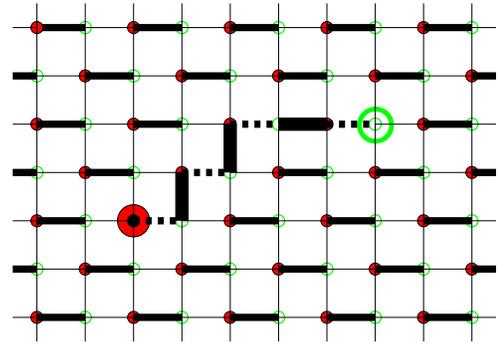}}
\vskip 0.2cm
\caption{``String'' mechanism for generating low-energy 
states in the band insulator phases.  Filled and open circles
represent $A$ and $B$ sites, respectively.  
Medium thick bonds represent a particular background dimer
pattern (i.e., bonds that are typically stronger), broken 
bonds from this set represent couplings that happened to be 
atypically weak in a given disorder realization, while the heavy 
thick bonds represent atypically strong couplings.
In the above figure, there is a pair of low-energy states 
associated with the end-points of the resulting string.
}
\label{string}
\end{figure}

In the dimerized band insulator phase, however, there is an 
insulating background to start with, and to produce such a pair 
one only needs to ``break'' the background dimer pattern along 
a string joining the two sites---i.e., of order $l$ specific events 
(see Fig.~\ref{string}), with the resulting occurrence rate 
$\sim\! e^{-c'_p l}$ high enough to give a power-law contribution 
to the low-energy density of states.  Our numerical RG and
exact diagonalization studies confirm that this is indeed what 
happens in such band insulator phases.  Furthermore, 
this ``Griffiths-string'' argument goes through for localized 
phases in any dimension.

\narrowtext

In the ``delocalized'' phase, on the other hand, we do not have 
such an insulating background.  However, the local properties of 
the ``isolating'' region need not be very ``sharp'', as was 
assumed in making the estimate for the probability of occurrence
$\Prob \!\sim\! e^{-c_p l^d}$.  Rather, there might
be some smooth ``textures'' of the random pairing (``dimer'') 
pattern producing such isolated sites (``defects''), and these may 
occur with much higher probability.

To be more specific, first consider the ``random-surface'' model
Eq.~(\ref{tPhi}) (i.e., the random vector potential model) 
in the limit of strong disorder $g \!\gg\! 1$.  The RG, being 
exact for the zero-energy wave function, 
preserves the random-surface character of the problem leaving
$\Phi(\br)$ unchanged: the randomness in the effective couplings 
that are generated by the RG rule~Eq.~(\ref{RGrule}) is of the same 
random-surface form~Eq.~(\ref{tPhi}) with the same $\Phi(\br)$.  
Thus, at each stage of the RG, the magnitude of the coupling 
between any two remaining sites $\alpha$ and $\beta$ is roughly
$\sim\! e^{\Phi(\alpha)-\Phi(\beta)}$.  Since the RG eliminates 
the stronger couplings first, an $A$-site $\alpha_{\min}$ 
that is a local minimum of $\Phi(\br)$ cannot be decimated 
out until the ``probing'' log-energy scale is large enough to
``see'' the next even lower minima:  
Indeed, before this happens, the other $A$-sites in the neighborhood
of $\alpha_{\min}$, having higher $\Phi$, are more strongly coupled 
to any available $B$-sites in the same neighborhood.  The same holds
for any $B$-site $\beta_{\rm max}$ that is a local maximum
of $\Phi(\br)$.  As a result, within our strong-randomness 
RG picture, the low-energy states of the problem
are associated with the local extrema of the surface
$\Phi(\br)$, while their effective log-energies are given by
the corresponding surface height differences.  In particular, 
in a finite system, the smallest energy scale is associated with 
the global extrema of $\Phi(\br)$.  The arguments leading to 
Eqs.~(\ref{intro:long:start})--(\ref{intro:zg:strong}) of the
Introduction now follow.

Thus, within our somewhat heuristic RG approach, ``isolated'' 
sites are produced often enough to result in a power-law 
contribution to the density of states, with the dynamical exponent 
$z \!\sim\! \sqrt{g}$ in the strong randomness regime.
Here, we are ignoring effect of running the RG 
transformation on the pure model itself---this will change 
the actual value of the exponent $z$, but not the 
$\sim\! \sqrt{g}$ disorder contribution for $g \!\gg\! 1$.  
For more precise bounds that complement these heuristics and lead 
to identical conclusions, see Sec.~\ref{sec:BOUNDS:longtd}.


In the general case, the heuristic RG arguments
Eqs.~(\ref{intro:gen:start})--(\ref{intro:gen:stop}) in the
Introduction now follow and yield the modified-Gade 
forms~Eq.~(\ref{rho:gade:x})~and~(\ref{length:gade:x}) with
$x=3/2$; more precise arguments leading to identical conclusions
are presented in Sec~\ref{sec:BOUNDS:general}.  However, 
it should be emphasized again that we have not found a clear-cut 
separation of the random vector potential and random mass parts 
of the disorder at the lattice level.
We therefore have to rely on the results of a different 
field-theoretic RG analysis\cite{CarOst,HwaDsf} when discussing 
how the two types of randomness flow and affect each other
(note that a similar somewhat heuristic argument was used in
Ref.~\onlinecite{DouGia} to obtain crossover scales in the vortex 
glass problem).

We conclude this discussion of the origin of low-energy states 
with some miscellaneous observations: 
Recall that the surface in the random vector potential case is also 
the logarithm of the pseudo-zero-energy-wave function 
$\Psi_A(\alpha) \!\sim\! e^{-\Phi(\alpha)}$.  
Given that the RG rule is an exact transformation for such a 
wave function in the general case as well,
the above analysis suggests that the $E\!\neq\!0$ low-energy 
states in the general problem are again related to the extremal 
properties of this $E\!=\!0$ wave function.  
It is also useful to note the predictions of this ``surface''
mechanism in other dimensions.  In one dimension, any disorder is 
of purely ``longitudinal'' (random-surface) type, and the Gaussian 
surface is more rough, with
\begin{equation}
|\ln E| \sim \Phi_{\max}(L) - \Phi_{\min}(L) \sim \sqrt{L} ~;
\end{equation}
this is precisely the Dyson critical scaling for the 1D chain.
For bipartite systems that are finite in all but one direction 
(such as ladders\cite{BroMudSimAlt,BroMudFur}), the DOS again has 
the Dyson form at special delocalized critical points in the phase 
diagram, since the corresponding surface ``looks'', at long 
wavelengths, like a 1D Gaussian surface.
On the other hand, in dimensions three and higher
the Gaussian surface is basically flat, and this mechanism 
cannot generate power-law contributions to the density of states.  
Much of what has been said above for the two-dimensional problem 
is therefore very specific to 2D.

\subsection{From low-energy states to optimized defects}
\label{sec:RG:struct}
We now establish a suggestive connection between the low-energy 
states in the bipartite hopping problem and monomers (defects) 
in the corresponding random dimer problem.  The dimer connection 
leads us to a heuristic ``dimer RG'' prescription for the low-energy 
states---this dimer RG approach provides us with a very useful 
physical picture that underlies the more precise analysis of 
Section~\ref{sec:BOUNDS}.

Consider running the above ``zero-energy'' RG in some specified 
order.  (The naive strong-randomness RG described earlier uniquely 
prescribes a particular order by picking the strongest effective 
bond at each step, but it is important to realize that in the 
absence of any rapid flows to stronger effective disorder, 
there is nothing ``sacred'' about this ordering, and other sensible 
choices are possible.)  We denote the $A$- and $B$-sites decimated 
out at a given stage of the RG as
\mbox{$\FA=\{a_1, \dots, a_n\} \subset A$} and
\mbox{$\FB=\{b_1, \dots, b_n\} \subset B$} respectively.
The effective coupling between any two remaining sites 
$\alpha$ and $\beta$ is given by
\begin{equation}
t'_{\alpha,\,\beta} = t_{\alpha,\,\beta} - 
\mt_{\alpha,\,\FB} \,\, \mt^{-1}_{\FA,\,\FB} \,\, 
\mt_{\FA,\,\beta} ~.
\label{tprime}
\end{equation}
Here, $\mt_{\FC, \FD}$ is a matrix of the original bonds joining 
two subsets $\FC \subset A$ and $\FD \subset B$ 
(thus, $\mt_{\alpha, \FB}$ is a row-vector, $\mt_{\FA, \beta}$ is 
a column-vector, and $\mt_{\FA,\FB}$ is an $n \times n$ matrix).  
Expression~(\ref{tprime}) follows readily if we note how the RG 
``solves'' for the zero-energy wave function by eliminating 
any reference to the decimated sites, and this holds for any order 
of the decimations.  Furthermore, it is easy to check that this 
may be rewritten as
\begin{equation}
t'_{\alpha,\,\beta} = 
\frac{ \det \mt_{\alpha + \FA,\, \beta + \FB} }
     { \det \mt_{\FA,\,\FB} } ~,
\label{tprime2}
\end{equation}
where $\alpha \!+\! \FA \equiv \{\alpha, a_1, \dots, a_n\}$, 
and similarly for $\beta \!+\! \FB$. 
It is this form that we find useful below.

Now, recall from Sec.~\ref{sec:DEFS:dimer} that for any two 
such sets $\FC$ and $\FD$ of equal cardinality
\begin{equation}
\det \mt_{\FC,\,\FD} = 
\sum_\cvr \sgn(\cvr) \prod_{\la\alpha\beta\ra \in \cvr} 
                          t_{\alpha\beta}~, 
\end{equation}
where the sum is over all complete dimer coverings of 
$\FC \cup \FD$, and $\sgn(\cvr)$ denotes the appropriate permutation
sign for matching $\cvr$.  Each term of the sum has a form 
\mbox{$\pm \exp\big( -\EE_d[\cvr]/T_d \big)$}, with the dimer
bond energies $\benrg_{\alpha\beta}$ defined from 
$|t_{\alpha\beta}| = \Omega_0 e^{-\benrg_{\alpha\beta}/T_d}$.
In the limit of infinite randomness, or equivalently, zero 
temperature of the dimer system $T_d=0$, one term---the ground state
dimer covering---dominates the whole sum; 
obviously, the permutation signs play no role in this limit.

Consider first this infinite-randomness limit.  The largest term in
$\det \mt_{\alpha + \FA,\, \beta + \FB}$
corresponds to the optimal (ground state) dimer covering of 
$\{\alpha \!+\! \FA\} \cup \{\beta \!+\! \FB\}$,
while the largest term in $\det \mt_{\FA,\,\FB}$ corresponds to
the optimal covering of $\FA \cup \FB$.  The two optimal coverings 
``differ'' by a ``string'' connecting the ``added'' sites 
$\alpha$ and $\beta$, while the effective coupling 
$t'_{\alpha,\beta}$ in this infinite-randomness limit, 
Eq.~(\ref{tprime2}), is determined precisely by ``propagating'' 
along the string
$t'_{\alpha,\beta} = t_1 t_3 \dots t_{2k+1}/(t_2 \dots t_{2k})$.
Note that this string is the optimal (i.e., corresponding to 
the largest possible $|t'|$) such path from $\alpha$ to $\beta$ 
utilizing the sites decimated out.

This suggests the following infinite-randomness ``dimer RG''
sequence:  At the $n$-th stage of such a dimer RG analysis, we find
the ground state configuration $\cvr^{(gs)}_n$ with $n$ dimers,
defining $\EEgs_n \!\equiv\! \EE_d[\cvr^{(gs)}_n]$.  The $2n$
sites covered by dimers at this stage are to be thought of as
having already been decimated.  At each step of this RG, 
precisely two sites are ``integrated out'' since the optimal 
$(n+1)$-dimer covering differs from the optimal $n$-dimer covering 
by two added sites (and the connecting string).  We now need 
to provide, for the random hopping problem, some notion of
the  bandwidth (i.e the value of the cutoff energy) 
corresponding to each stage.  
A natural choice is to associate the effective hopping amplitude 
$|t'| = \Omega_0 \exp\big[-(\EEgs_{n+1}-\EEgs_n)/T_d \big]$ 
with the two sites ``eliminated'' in going from stage $n$ to 
$n + 1$---this then specifies the value of the cutoff in
the hopping problem at this RG step.
The monotonicity property 
$\EEgs_{n+1}-\EEgs_n \geq  \EEgs_n-\EEgs_{n-1}$
guarantees that the cutoff indeed decreases monotonically
as the RG proceeds.  Note, however, that the ``pairings'' of 
the sites decimated out are somewhat subtle; these do not stay 
rigid as the RG proceeds, and can change each time we decimate 
more sites.  Furthermore, there is no simple relationship in 
general between the sequence of decimations within this dimer RG, 
and that in the naive strong-disorder RG sequence described earlier.

Of course, this can all be reformulated starting from the 
ground state complete dimer covering (with $N=L^2/2$ dimers) of 
the whole lattice, and then adding pairs of monomers (defect pairs) 
into the system in the most optimal way.  In particular, 
within this heuristic picture, the lowest 
energy state in the hopping problem corresponds precisely to 
the first such pair, with the corresponding estimate
\begin{equation}
|t'|^{\rm (dimerRG)}_{\min} 
\equiv \Omega_0 \exp\Big( \frac{\Eoptdef}{T_d} \Big) ~;
\label{Emin:dimerRG}
\end{equation}
here $\Eoptdef \equiv \EEgs_{N-1}-\EEgs_N$ is the optimized 
(lowest energy) defect pair energy, and all of the foregoing
assumes the system has complete coverings 
to start with.\cite{GraphZeroes}

Returning to the finite-randomness case, there are several issues 
that stand out when trying to relate the low-energy states of 
the hopping problem to the monomers in the corresponding 
finite-temperature random dimer system.  The determinant 
$\det \mt_{\FA,\,\FB}$ and the dimer partition function 
$Z_d[\FA, \FB]$ still contain the same (in absolute value) 
terms, but now the dimer ground state by itself does not determine 
the corresponding sums, so the permutation signs become 
important.  Even if we use the dimer partition function 
only to bound the absolute value of the determinant, we are still 
faced with the dimer problem at finite temperature, which now
has significant entropic contributions, since local moves from 
the ground state cost only finite energy.  In this situation, 
we can no longer think solely in terms of the ground state pairing 
when attempting to estimate $Z_d[\FA, \FB]$.
  
However, we are primarily interested in the ratios of
determinants like $\det \mt_{A,\,B}$ and 
$\det \mt_{A-\alpha,\,B-\beta}$, i.e., the difference in 
the corresponding free energies of the dimer system without defects 
and with two defects.  Since a large part of the entropic terms 
in the two systems comes from the same local moves, the ground state
energy difference of the two may still be a relevant object.
Put another way, we need to consider the 
{\it effective thermodynamics} of the defects themselves---i.e., 
with all free energies measured relative to the defect-free 
system.  As we shall see within a more precise formulation
in Sec.~\ref{sec:BOUNDS:general}, this effective thermodynamics 
is indeed ground-state dominated, allowing us to make fairly 
precise statements for finite values of bare disorder as well.

\section{``Variational'' bounds on the states near the band center}
\label{sec:BOUNDS}
Motivated by these heuristic RG considerations, we now establish 
more rigorous bounds on the dynamical scaling in the system.
Consider the bipartite Hamiltonian~(\ref{H}) in a finite sample 
and assume that there are no zero-energy states and no edge states;
a concrete realization would be some regular lattice block with 
some nice boundary conditions, e.g., the $\pi$-flux model
on an $L \!\times\! L$ square with free bc and even $L$.  
The smallest positive energy 
$E_{\min} \equiv E^+_{\min}$ of $\hat H$ is then given as
\begin{equation}
E_{\min} = \frac{1}{\|\mt^{-1}_{AB}\|} ~.
\label{Emin}
\end{equation}
Here, $\| \hat C \|$ is the matrix norm of an operator $\hat C$, 
and satisfies the inequalities
\begin{equation}
\max_{i,j} |C_{ij}| \leq \|\hat C\| \leq 
\sqrt{\sum_{i,j} |C_{ij}|^2} \leq \sum_{i,j} |C_{ij}| ~.
\label{Cnorm}
\end{equation}
Expressions~(\ref{Emin})~and~(\ref{Cnorm}) formalize our
earlier, more heuristic association of low-energy states with 
optimized defects in the dimer problem.  Indeed, we have 
\begin{equation}
(\mt^{-1}_{AB})_{\beta\alpha} = \pm
\frac{\det\mt_{A-\alpha,\,B-\beta}}{\det\mt_{AB}} ~,
\label{tinv}
\end{equation}
which is precisely the previously encountered ratio of the 
determinants, and the above inequalities  suggest that the two 
``defect'' sites $\alpha$ and $\beta$ should be chosen in some 
optimal way.

Below, we use these inequalities to obtain, most importantly, 
{\it lower bounds} for the smallest positive energy.\cite{boundrg}

\subsection{The random vector potential model}
\label{sec:BOUNDS:longtd}
Applying the above bounds to the ``random-surface'' model 
Eq.~(\ref{tPhi}), we obtain, in particular,
\begin{equation}
\|\mt^{-1}_{AB}\|^2 \leq  \sum_{\alpha,\beta} 
e^{2\Phi(\beta)-2\Phi(\alpha)} |G_0(\beta,\alpha)|^2,
\label{ZGsum}
\end{equation}
where $G_0(\beta, \alpha) \!\equiv\! (\mt^{-1}_0)_{\beta\alpha} 
\!\equiv\! \la\beta | \hat H^{-1}_0 | \alpha \ra$ is the $E\!=\!0$ 
Green's function (propagator) for the pure lattice problem.  
The propagator $G_0$ can be calculated explicitly for each 
particular pure system.  In the following, we consider specifically 
the $\pi$-flux lattice model with free bc; the lattice propagator 
(Dirac fermions) can be found, e.g., in Ref.~\onlinecite{FisSte}, 
but here we need only note that $G_0(\br_1, \br_2)$ is bounded 
and behaves as 
$|G_0(\br_1, \br_2)| \!\sim\! |\br_1-\br_2|^{-1}$ 
for large $|\br_1-\br_2| \!\gg\! 1$.  A simple (completely rigorous) 
bound $\|\mt^{-1}_{AB}\| \!\leq\!  \const \times \FZ (L)$
immediately follows from the boundedness of $G_0$.
Here, $\FZ (L)$ is the partition function for 
a classical particle living in a reduced potential 
\mbox{$V(\br)/(k_B T)=2\Phi(\br)$}---see Appendix~\ref{app:2dGAUSS}, 
where we also summarize other relevant properties of such 
2D Gaussian surfaces.  Furthermore, it is actually possible to 
provide stronger bounds---below we discuss this separately for the 
{\it strong-randomness} ($g > g_c \equiv 2\pi$) and 
{\it weak-randomness} ($g < g_c$) regimes
(see Appendix~\ref{app:2dGAUSS} for the significance of $g_c$ and
the distinction between the two regimes in the random-surface 
problem).

In the strong-randomness regime,
we expect that the sum~(\ref{ZGsum}) is dominated by the $\beta$'s 
near the maxima of $\Phi(\br)$ and the $\alpha$'s near the minima 
of $\Phi(\br)$ [analogous to Eq.~(\ref{ZsumMax}) for 
$\FZ (L)$].  We also expect that these extrema are separated by 
distances of order $L$ (the precise statement would be that 
the distance between the global maximum and minimum of $\Phi(\br)$ 
is $L$ times a random number of order one).  This suggests that 
\begin{equation}
\|\mt^{-1}_{AB}\| \leq 
\num \times \frac{e^{\Phi_{\max}(L)-\Phi_{\min}(L)}}{L}
\sim \frac{\FZ (L)}{L}~.
\label{tupper}
\end{equation}
The expected uncertainty in the rhs of this expression from one 
disorder realization to another is simply a random $O(1)$ numerical 
factor $\num$.  [Note that even though the above estimate of the
sum~(\ref{ZGsum}) does not constitute a rigorous proof, it is 
fairly robust---moreover, a more formal proof is likely possible 
along the lines of Refs.~\onlinecite{CasChaFraGolMud,CarDou01}.]  
Since the sum in the estimate is essentially dominated by few 
terms, this also provides a lower bound for $\|\mt^{-1}_{AB}\|$,
that differs from the upper bound~(\ref{tupper}) only by an
$O(1)$ numerical factor:
\begin{eqnarray*}
\|\mt^{-1}_{AB}\| 
\geq  \max_{\alpha,\beta} | (\mt^{-1}_{AB})_{\beta\alpha} |
=  \const \times \frac{e^{\Phi_{\max}(L)-\Phi_{\min}(L)}}{L} ~;
\end{eqnarray*}
the bound is rigorous (and completely general) since the distance
between the global extrema cannot exceed the sample size.
Thus, the smallest positive energy is expected to scale with 
the system size as
\begin{equation}
E_{\min} = \num \times \frac{L}{\FZ (L)} \sim
\frac{ (\ln L)^{ \frac{3}{2}\sqrt{\frac{g}{g_c}} } }
     { L^{4\sqrt{\frac{g}{g_c}}-1} } ~.
\label{Emin:wLogCorr}
\end{equation}
In particular, for the dynamical exponent we obtain
\begin{equation}
z=4\sqrt{\frac{g}{g_c}}-1~, \quad g>g_c ~.
\label{zg:strong}
\end{equation}
This is our central result; pictorially, the surface $\Phi(\br)$
simply does not allow the low-energy states to fall beyond its 
extrema.  
[In Eq.~(\ref{Emin:wLogCorr}) we have also included the logarithmic 
correction from~Ref.~\onlinecite{CarDou01} to show that the minimal 
energy is somewhat larger than expected from the simple-$z$ 
power-law; note, however, that the actual form of such 
log-corrections can, in general, depend on the boundary 
conditions used.\cite{CarDou01}]

In the weak-randomness regime, the bound
Eq.~(\ref{ZGsum}) is not as precise. 
For instance, in the pure-system ($g=0$) case, we obtain
\begin{equation}
\sum_{\alpha,\beta} |G_0(\beta,\alpha)|^2 
= {\mathrm Tr}\, \hat H_0^{-2} \sim L^2 \ln L ~,
\end{equation}
i.e., $E_{\min} \geq L^{-1} (\ln L)^{-1/2}$, which gives 
the correct exponent $z=1$, but is certainly an underestimate.  
Similarly, in the $g \neq 0$ case, we can prove that for any 
$\eta > 0$
\begin{equation}
\|\mt^{-1}_{AB}\| \leq \const \times L^{1+2g/g_c +\eta}
\end{equation}
with probability one in the limit $L \to \infty$.  The proof 
(following the lines of Ref.~\onlinecite{CasChaFraGolMud})
applies the inequality $\Prob (X \geq K) \leq \la X \ra /K$,
valid for any nonnegative random variable $X$, to the particular
$X$ equal to the rhs of Eq.~(\ref{ZGsum}).  Thus, the dynamical
exponent $z$ obeys the following bound quite generally:
\begin{equation}
z \leq 1+\frac{2g}{g_c} ~.
\label{weak:zleq}
\end{equation}
Note that this result is valid in the strong disorder regime 
as well---however, in this case an individual realization 
essentially never samples the tails of the distributions that 
determined the average $\la X \ra$, and our earlier treatment 
leading to Eq.~(\ref{zg:strong}) is more appropriate and provides 
a much sharper bound.
 
For weak randomness $g<g_c$, we can also give a plausible argument 
for the opposite inequality $z \geq 1+2g/g_c$.  It suffices to 
provide an upper bound for $E_{\min}$, and a good trial wave 
function for this seems to be 
\begin{equation}
\vec\psi_B = \mt^{-1}_{AB} e^{-\Phi(\vec A)} ~,
\end{equation} 
with 
$E_{\min} \leq \|\mt_{AB} \vec\psi_B\| / \|\vec\psi_B\|$
bounded by
\begin{equation}
E_{\min}^2 \leq 
\frac{ \sum_\alpha e^{-2\Phi(\alpha)} }
     { \sum_\beta e^{2\Phi(\beta)} 
      \Big(\sum_\alpha G_0(\beta,\alpha) e^{-2\Phi(\alpha)}
      \Big)^2 } ~.
\end{equation}
To translate this into an actual lower bound on $z$, we first
estimate the sum over $\alpha$ sites in the denominator as
\begin{equation}
\Big| \sum_\alpha G_0(\beta,\alpha) e^{-2\Phi(\alpha)} \Big|
\sim  \sum_\alpha | G_0(\beta,\alpha) | e^{-2\Phi(\alpha)} ~.
\end{equation}
This estimate is clearly reasonable, since
$G_0(\br_1, \br_2)$ is some fixed function, while 
the sums involving $e^{-2\Phi(\alpha)}$ over macro-subregions 
(with $G_0\!>\!0$ and $G_0\!<\!0$) always have $O(1)$ random
numerical factors [see, e.g., expressions for $\FZ(L)$ in Appendix].
Now, using $|G_0| \geq L^{-1}$ we obtain 
$E_{\min} \leq \num \times L /\FZ (L)$, which gives the desired
coinciding lower bound for $z$ in the weak-disorder regime 
(note that this also provides a similar coinciding lower bound 
in the strong-disorder regime, and thus reproduces our result 
for $z$ in this case).

\narrowtext
\begin{figure}
\epsfxsize=\columnwidth
\centerline{\epsffile{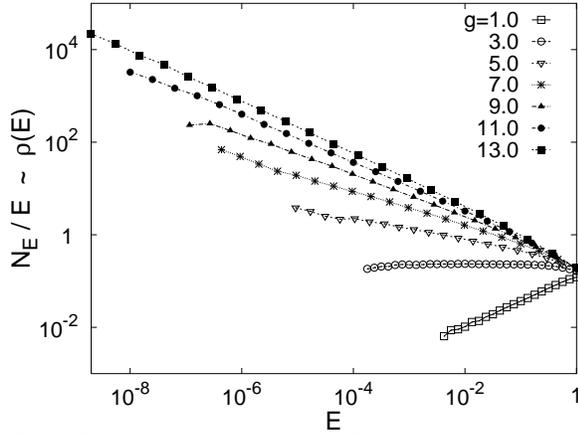}}
\caption{Density of states in the $\pi$-flux model with the
``random-surface'' disorder [see Eq.~(\ref{tPhi})] calculated for 
$192 \times 192$ system with open boundary conditions, for 
varying disorder strength $g$.  The data is averaged over $10$ 
disorder realizations.  Note that the plots have some 
curvature for strong disorder---this is probably because of 
the logarithmic corrections to the power-laws in this regime.
}
\label{PiFluxVlong}
\end{figure}

\narrowtext
\begin{figure}
\epsfxsize=\columnwidth
\centerline{\epsffile{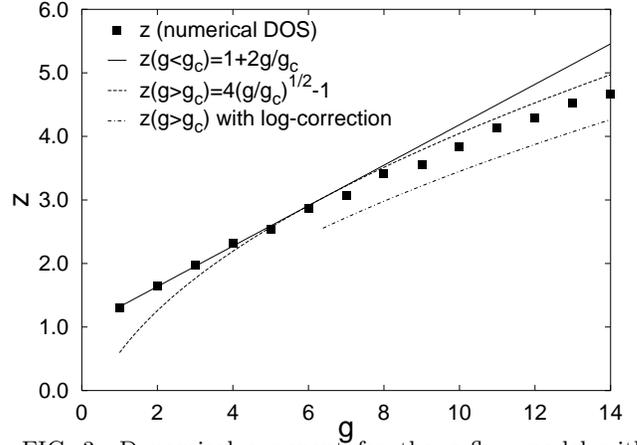}}
\caption{Dynamical exponent for the $\pi$-flux model with 
the random-surface disorder extracted from the
data of Fig.~\ref{PiFluxVlong}.  The additional lines
show theoretical predictions in the two regimes (with $g_c=2\pi)$), 
and also a rough estimate 
$z_{\rm eff}(L) = z(\infty) - (3/2)\,\sqrt{g/g_c}\,
(\ln\ln L)/(\ln L)$ of the log-correction in the
strong-disorder regime for the specific system size studied.
}
\label{zPiFluxV}
\end{figure}

We conclude with two comments on the results obtained here.
Firstly, although we cannot give a proof, we expect 
that the expression for $z$ in the weak-disorder regime,
\begin{equation}
z = 1+\frac{2g}{g_c}\,,\quad g<g_c ~,
\label{zg:weak}
\end{equation}
is exact without any logarithmic corrections.
Our expectation is bolstered by the fact that Eq.~(\ref{zg:weak})
coincides with the original result of the weak-disorder
replica analysis of Ref.~\onlinecite{LudMpafShaGri} 
[which is known to give exact results also for the static 
quantities of Appendix~\ref{app:2dGAUSS} in the regime $g<g_c$].  
Moreover, the two expressions for $z(g)$, 
Eqs.~(\ref{zg:strong})~and~(\ref{zg:weak}), join continuously 
and smoothly at $g_c$ [and we expect that
$E_{\min} = \num \times L /\FZ (L)$ is valid for any disorder
strength].

Secondly, note that the above results are, strictly speaking,
valid only for the exponent $z_1$ that determines the scaling of 
the smallest gap in the system.  However, this is not expected 
to be a serious issue:  Thus, having the upper bounds on 
$E_{\min}(L)$, we are guaranteed a bulk density of states with 
the DOS dynamical exponent $z_{\rm DOS}$ at least as large.
Moreover, we expect quite generally that there is only one dynamical 
exponent in the system $z_{\rm DOS} \equiv z_1\equiv z$.  
Some numerical checks for the bulk density of states are shown 
in Figs.~\ref{PiFluxVlong}~and~\ref{zPiFluxV}, and these are 
in good agreement with the above expectations.
(See also Ref.~\onlinecite{CasDou} studying essentially 
identical dynamical freezing phenomenon for random walks in 
random log-correlated potentials in one and two dimensions.)

\subsection{Scaling of the lowest energy in the general case}
\label{sec:BOUNDS:general}
In general, we can always bound the numerator in~Eq.~(\ref{tinv}) 
by the corresponding dimer partition function with two {\em fixed} 
monomers $\alpha$ and $\beta$
as $|\det \mt_{A-\alpha,B-\beta}| \leq Z_d[A-\alpha,B-\beta]$.
We now specialize to the $\pi$-flux model with free bc (or any 
other model with a direct connection to dimers), so that for 
the lattice with no defects we have 
$|\det \mt_{AB}| = Z_d[A,B] \equiv Z_N$, with $N=L^2/2$ 
the total number of dimers.  In this case, we do not have to worry 
about the denominator at all, obtaining, e.g.,
\begin{equation}
\|\mt^{-1}_{AB}\| \leq 
\frac{\sum_{\alpha\beta} Z_d[A-\alpha,B-\beta]}{Z_d[A,B]} = 
\frac{Z_{N-1}}{Z_N} ~,
\label{Zdefect}
\end{equation}
where $Z_{N-1}$ is the partition function of the dimer problem
with two {\em free} defects allowed to be anywhere on the lattice.

Now, this ratio of partition functions describes the effective 
finite temperature thermodynamics of a pair of defects 
(i.e., with all contributions to the free energy measured
relative to that of the dimer system with no defects).
The analysis of Ref.~\onlinecite{ZenLeaDsf} suggests that 
this effective finite-temperature thermodynamics is equivalent
to the thermodynamics of a pair of classical charged particles
(with opposite charges for $A$ and $B$ monomers) in a random 
``electrostatic'' potential $\upsilon(\br)$ with average correlations
\begin{equation}
\overline{[\upsilon(\br)-\upsilon(\br')]^2} 
\sim \ln^2 \!|\br-\br'| ~.
\label{vcor}
\end{equation}
The extremal properties of this random surface $\upsilon(\br)$ 
have also been characterized in Ref.~\onlinecite{ZenLeaDsf}.  
In a finite box of linear size $L$, we have for the average 
and the standard deviation of the distribution of the global extrema
\begin{equation}
\overline{\upsilon_{\max}} \sim (\ln L)^{3/2}, \qquad 
\sigma(\upsilon_{\max}) \sim (\ln L)^{1/2} ~,
\label{vmax}
\end{equation}
Reasoning as in Ref.~\onlinecite{CarDou01}, we see that these 
extremal properties immediately imply
\begin{equation}
\frac{Z_{N-1}}{Z_N} \leq 
L^4 \exp \left[ \frac{\upsilon_{\max}-\upsilon_{\min}}{T_d} 
              \right] 
\leq \exp \left[ \const \times (\ln L)^{3/2} \right] ~,
\label{ratZN}
\end{equation}
and correspondingly for the lower bound on the smallest positive
energy
\begin{equation}
\ln E_{\min} \geq  -\const \times (\ln L)^{3/2} ~.
\label{Emin:gen}
\end{equation} 
In the above, making the positive number ``$\const$'' slightly 
larger takes care of any prefactors $L^r$ and any uncertainty
$o((\ln L)^{3/2})$ in $\upsilon_{\max}$.  This result is a simple 
manifestation of {\em ground state dominance}\cite{CarDou01} for
the finite-temperature partition sum of particles moving in the 
random potential $\upsilon(\br)$ with correlations~Eq.~(\ref{vcor}).
The above crude estimates are good enough to reach the main 
conclusion~Eq.~(\ref{Emin:gen}) precisely because of this, 
even though the ground state dominance is a very weak one. 
Since a few defect positions effectively determine 
the above estimates, we expect that Eq.~(\ref{Emin:gen}) 
provides not only the lower bound but the actual leading 
term in the logarithm of the smallest positive energy 
in the system.  This immediately leads to the modified-Gade
scaling proposed in the Introduction---our analysis here
thus provides strong evidence in support of the RG arguments 
outlined in the Introduction.

\vskip 0.1cm
To get some idea about the subleading terms and, 
in particular, the relationship between the actual lowest gap,
the intermediate bound Eq.~(\ref{Zdefect}) obtained from the 
effective thermodynamics of a free defect pair, and the heuristic 
dimer RG ``estimate'' Eq.~(\ref{Emin:dimerRG}) from the optimized
defect pair, it is instructive to consider, from this perspective, 
the random vector potential model in the strong-randomness regime
$g \!\gg\! 1$.

Recall first our earlier direct estimate of $\| \mt_{AB}^{-1} \|$,
i.e., the lhs of Eq.~(\ref{Zdefect}).
Using the Dirac fermion propagator for $\mt_0^{-1}$, we have
\begin{equation}
|(\mt_{AB}^{-1})_{\beta\alpha}| 
\approx \frac{\const}{|\br_\beta-\br_\alpha|}
\times  e^{\Phi(\beta) - \Phi(\alpha)} ~.
\label{dett}
\end{equation}
Invoking the ground state dominance in the strong-randomness 
regime, we obtain
$\| \mt_{AB}^{-1} \| \sim L^{-1} e^{\Phi_{\max}-\Phi_{\min}}$.

Consider now the relative partition sum $Z_{N-1}/Z_N$ 
for a free defect pair, i.e., the rhs of Eq.~(\ref{Zdefect}).  
In this ``random-surface'' case, for a given fixed configuration 
of monomers, any dimer covering has the same energy, and 
the partition function is essentially purely configurational 
(entropic).  Thus, for two fixed defects $\alpha$ and $\beta$, 
we have
\begin{equation}
\frac{ Z_d [A-\alpha,B-\beta] }{ Z_d [A,B] }
\approx \frac{\const}{|\br_\beta - \br_\alpha|^{1/2}} 
\times e^{\Phi(\beta) - \Phi(\alpha)} ~,
\label{Zt}
\end{equation}
since the relative number of dimer configurations decays as 
$|\br_\beta - \br_\alpha|^{-1/2}$ (Ref.~\onlinecite{FisSte}) 
with increasing separation between the monomers.
Taking into account the ground state dominance in the 
strong-randomness regime, we then estimate
$Z_{N-1}/Z_N \sim L^{-1/2} e^{\Phi_{\max}-\Phi_{\min}}$.   
Thus, the rhs of the inequality~(\ref{Zdefect}) overestimates 
the lhs by a factor of $L^{1/2}$, and the difference comes about 
precisely because of the permutation signs in the expansion of 
$\det\mt_{A-\alpha,B-\beta}$ 
[compare Eqs.~(\ref{dett})~and~(\ref{Zt})].

Finally, the defect pair energy for the optimized 
positions is simply $\Eoptdef/T_d = -(\Phi_{\max}-\Phi_{\min})$,
and we see that $Z_{N-1}/Z_N \sim  L^{-1/2} \exp(-\Eoptdef/T_d)$;
the factor $L^{-1/2}$ here represents the entropic cost for
introducing two essentially fixed monomers into the dimer system.

Putting everything together, we may thus write 
$E_{\min} \!\approx\! \num \times L \exp(\Eoptdef/T_d)$.
Thus, the dimer RG expression~(\ref{Emin:dimerRG}) 
{\it underestimates} the smallest positive energy by a factor 
$L^{-1}$ coming from the fermionic sign and dimer entropic origins.

Returning to the general case, we expect the intermediate bounds, 
as well as the dimer RG estimate from the optimized defects, 
to underestimate the smallest positive energy by some similar 
$L^{-r}$ factors. 
In our numerical tests (see below), 
we indeed find that the dimer RG provides a strong lower bound 
for $E_{\min}$.  However, we cannot be more precise in the general
case because of the following difficulty:
Unlike the random vector potential case, 
we cannot disentangle the defect energetics from the dimer 
configurational combinatorics.  More explicitly, the results of 
Ref.~\onlinecite{ZenLeaDsf} suggest that a fixed pair of defects 
will have an average energy $O(+\ln |\br_\alpha - \br_\beta|)$ 
because the system constrained by the presence of the defects 
cannot take as much advantage of the low energy bonds, 
and $\Eoptdef$ will have a similar $O(+\ln L)$ contribution
in addition to the dominant $O(-(\ln L)^{3/2})$ term
coming from $\upsilon_{\min}-\upsilon_{\max}$.


\section{Numerical tests}
\label{sec:NUMERICS}
We have performed extensive numerical tests using exact
diagonalization and numerical RG.  The results are consistent 
with the generalized-Gade scaling Eq.~(\ref{length:gade:x}) 
in the delocalized phase.  Although direct diagonalization methods 
do not allow us to unambiguously distinguish the different exponents 
$x$ in the generalized-Gade scaling, indirect methods provide strong
indications in favor of $x=3/2$, different from the original Gade 
prediction of $x=2$.  In the localized phase, the numerical results 
are more conclusive, and all methods clearly point to a power-law 
density of states (with nonuniversal dynamical exponent) 
generated by the ``string'' Griffiths mechanism discussed earlier.

\subsection{Illustration}
\label{sec:NUMERICS:ensample}
\begin{figure}[t]
\epsfxsize=\columnwidth
\centerline{\epsffile{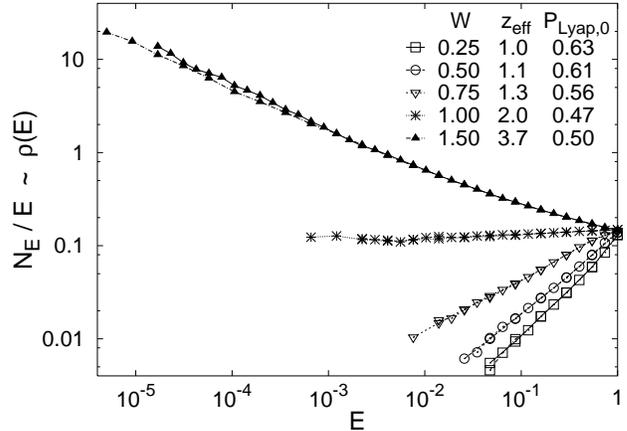}}
\vskip 0.1cm
\caption{Illustration of the ``apparently nonuniversal'' 
behavior of the density states (normalized per lattice site).  
The system is a $192 \times 192$ honeycomb lattice with hopping 
elements chosen randomly from a uniform distribution in $[1-W,1+W]$;
free boundary conditions are used in one direction and periodic bc 
in the other [see Fig.~\ref{brickwall}(a)].
The data is averaged over $10$ disorder realizations; 
the lowest energy shown in each case corresponds roughly to 
the second-lowest energy state in the finite sample.  
Similar data for a smaller $96 \times 96$ system is also shown;  
from such comparisons, we conclude that the density of states 
plotted is indeed the bulk density of states.  
The effective dynamical exponent $z_{\mathrm eff}$ is obtained by 
fitting the integrated density of states $N_E$ to the form 
$N_E \sim E^{d/z}$ over the range $2 \!<\! L^2\,N_E \!<\! 200$ 
in each case.  Direction-averaged Lyapunov spectrum density 
$\overline\rlyapz$ is calculated from numerical transfer matrix 
studies and provides a rough idea on the conducting properties 
of the system.
}
\label{HNWscan}
\end{figure}

We first demonstrate some difficulties that plague direct 
diagonalization studies.  In this context, it is important 
to note that a strong Gade-like divergence
in the low-energy density of states has never been observed in
previous numerical studies.\cite{Eva86,MorHat}  In our direct 
diagonalization studies, we typically see power-law diverging or 
even vanishing density of states with nonuniversal exponents.
This is illustrated in Fig.~\ref{HNWscan} where we show 
the calculated\cite{numDOS:short,pbc1:short} density of states 
$\rho(E)$ for the honeycomb lattice with hopping amplitudes chosen 
uniformly from $[1-W,1+W]$.  For $W < 1$, on the length 
scales studied, the density of states is apparently power-law 
vanishing, with the observed effective exponent $z_{\rm eff}$ that 
depends on the strength of disorder $W$.  It is simply impossible
from the data to guess that there may be something else happening 
at still lower energy scales!  For $W=1.0$ the density of states 
is almost constant, with a slight hint on some weak divergence at 
still lower energies observed for the largest system studied.  
For $W>1.0$ (the honeycomb lattice is still statistically isotropic 
in spite of the negative hopping amplitudes allowed, 
but the direct dimer connection is lost in this case), 
a power-law divergence is observed, and in this case, we also see 
some curvature in the log-log plot of $\rho(E)$ vs $E$ suggestive 
of an even stronger divergence at still lower energy; 
however, from these numerical results it is not possible to say 
anything more about the nature of the true low-energy singularity 
in the DOS.

A partial resolution of this discrepancy between direct numerics 
and theory lies in the fact that the asymptotic regime, in which the
form~Eq.~(\ref{rho:gade:x}) is expected to hold, happens to
be quite inaccessible for the exact diagonalization studies, 
given the computational restriction on the system size. 
Taking the original results of Ref.~\onlinecite{Gade} at face 
value for the moment, we see that the asymptotic 
form is valid only below a crossover scale\cite{myEcross}
\begin{equation}
E_{\rm cross} \sim \Omega_0 e^{-16 \pi^2 y^2 \conduct^2} ~,
\label{Ecross}
\end{equation}
where $\Omega_0$ is some bare energy, $\conduct$ is the 
dimensionless conductivity of the 2D system, and $y$ is some 
dimensionless parameter that is greater than one.
The conductivity $\conduct$ that enters here can be
estimated as $\conduct \sim \rlyapz$ with some $O(1)$ prefactor; 
here $\rlyapz$ is the relevant Lyapunov spectrum density that 
can be obtained from a numerical transfer matrix analysis
(see the next subsection).  In Fig.~\ref{HNWscan}, we have 
therefore also listed the corresponding values of $\rlyapz$ 
obtained for our system.  The important thing to notice is that
for the typically studied systems, the estimated $\conduct$ is of
order one and  can vary somewhat.  To get some feeling for the 
numbers involved, we set $y=1$ and check how 
$E_{\rm cross}$ changes as we vary $\conduct$.  
For the three values $\conduct=1.0$, $0.5$, and $0.2$, we get 
$E_{\rm cross}/\Omega_0 = 10^{-69}$, $10^{-17}$, and $10^{-3}$
respectively!  For energies above the crossover energy,
Gade predicts an effective power-law density of states, and this is 
probably all that has been observed in the previous numerical 
studies (but see the very recent Ref.~\onlinecite{RyuHat}).

Furthermore, the modified-Gade form Eq.~(\ref{rho:gade:x}) 
with $x=3/2$ proposed here is a weaker divergence than 
the earlier prediction of Gade.  This is probably another reason 
why it has proved difficult to observe anything other than 
power-law singularities in the direct diagonalization studies.
Of course, we expect the actual crossovers to be correspondingly 
more complicated,\cite{DouGia} but estimates similar to the
ones obtained from Gade's original work are still expected to hold.
The upshot of all this is that one should therefore be very cautious 
when looking for the asymptotic behavior in two dimensions.

Finally, note that the foregoing suggests a way to bring the
crossover energy up into the range accessible to direct numerical 
studies---the idea is to make the system look less delocalized, 
and this is what we turn to shortly.

\subsection{The choice of the system}
\label{sec:NUMERICS:choice}

\begin{figure}
\epsfxsize=2.8in
\centerline{\epsffile{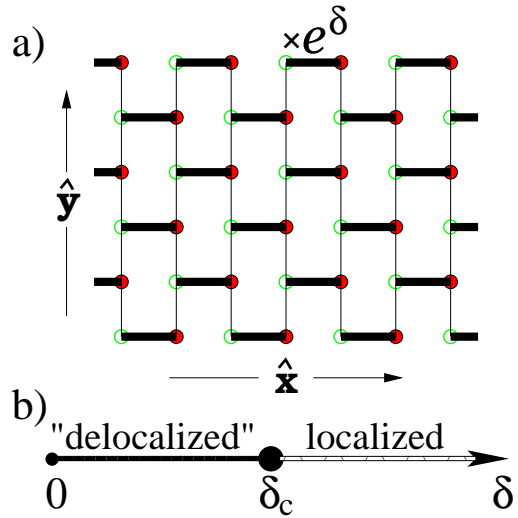}}
\caption{a) The brickwall lattice system we study numerically.
Different groups of hopping amplitudes are chosen from different
uniform distributions $[0,J]$, with $J=1$ for thin lines
(regular bonds) and $J=e^\delta$ for dark thick lines
(dimerized bonds).
b) Phase diagram of this system from transfer matrix analysis; 
$\delta_c \approx 1.432$.
}
\label{brickwall}
\vskip 0.15cm
\end{figure}

In the following, we present our numerical results only 
for one specific system---the brickwall (honeycomb) lattice
of Fig.~\ref{brickwall}(a).  Nearest-neighbor hopping 
amplitudes $t_e$ are taken randomly and independently from 
uniform distribution $[0,J_e]$.  When $J_e \equiv 1$ for all 
bonds of the lattice (i.e., all $t_e$'s are taken from the same 
distribution and the underlying honeycomb lattice is statistically
isotropic) the system is not exponentially localized at $E=0$, 
as can be checked by direct transfer matrix analysis.  
By allowing the horizontal bonds to be stronger on average, 
$J_e = e^\delta$ for the dark thick bonds in Fig.~\ref{brickwall}(a) 
(corresponding to an anisotropic honeycomb lattice), we can drive 
the system into a localized state for strong enough 
$\delta > \delta_c$, but the system remains ``delocalized'' 
for weak $\delta<\delta_c$ 
[see Fig.~\ref{brickwall}(b) for the phase diagram].

The rationale behind our specific choice of system is as follows: 
We want to perform all numerics on the same system, so that 
we can compare results of different approaches.  It is therefore 
best to use a system that has the precise connection 
$\det \mt_{AB}=Z_d[A,B]$ with the corresponding random dimer 
problem, so that we can further check our arguments 
of Sec.~\ref{sec:BOUNDS:general}.  
Moreover, we want a system whose statistically isotropic 
version (i.e., without any enforced dimerization in values of
the hopping amplitudes) shows an appreciable singularity in the 
low-energy DOS for accessible system sizes [this last requirement 
rules out the $\pi$-flux lattice---the largest $z_{\rm eff}$ that 
we can observe in this system with order one bare randomness is 
$z_{\rm eff} (L\!\!=\!\!192) \cong 1.5$]. 

Of course, the fact that our lattice has the precise dimer 
connection does not mean that the results presented below are 
non-generic.  We also studied the honeycomb and square lattices 
with random bonds of any sign and obtained qualitatively similar 
results.  In these systems with disorder strength of order one, 
we actually observe a stronger effective divergence in $\rho(E)$, 
which can be more readily checked against the precise functional 
form, but only through multi-parameter fits, which are not entirely 
convincing.  Since we can alternatively bring the crossover energy 
up into the accessible range by increasing $\delta$ and 
simultaneously preserving the dimer connection, this is what we 
choose to do, in part because the more precise ``dimer'' arguments 
provide us with the strongest evidence in favor of the proposed 
modified-Gade scaling.  However, we do expect that the 
suggestive strong-randomness association of the low-energy states 
in the hopping problem with the optimized defects in the dimer 
problem is likely still valid in some effective sense even when 
the fermionic problem does not have the precise $E=0$ dimer 
connection.

\subsection{Transfer matrix analysis}
\label{sec:NUMERICS:lyap}
We now summarize the transfer matrix analysis that gives us
the phase diagram Fig.~\ref{brickwall}(b), and also explain
our procedure for estimating the conductivity $\conduct$.
The general setting is as follows:\cite{ChaBer,MacKra}
Consider transferring the wave function along a strip of 
transverse size $N \equiv L_\perp$.  The transfer matrix is a 
$2N \times 2N$ matrix, and we calculate the Lyapunov spectrum 
of the corresponding random matrix product.  The $2N$ Lyapunov 
exponents $\nu_i$ come in $\pm$ pairs, and we focus on 
the positive half of the Lyapunov spectrum
\mbox{$\nu_1 \geq \dots \geq \nu_N \geq 0$}.
The largest localization length $\xi_{\max}$ in the strip is 
obtained as the inverse of the smallest (in absolute value)
Lyapunov exponent, $\xi_{\max} \equiv 1/\nu_N$.  If $\xi_{\max}(N)$
is found to increase linearly with $N$, the system is said to be 
delocalized (more precisely, the 2D system is not exponentially 
localized).  If, on the other hand,  
\mbox{$\xi_{\max}(N\to \infty)$} saturates to a finite value, 
the 2D system is localized.  Alternatively,\cite{ChaBer} 
the system is delocalized if there is a finite density of 
Lyapunov exponents
\begin{equation}
\rlyap(\nu) \equiv 
\lim_{N \to \infty} N^{-1} \sum_i \delta(\nu-\nu_i)
\end{equation}
at $\nu=0$; similarly, the system is localized if there is 
a gap in the Lyapunov spectrum around $\nu=0$.
Moreover, in the delocalized state, the Lyapunov spectrum density
$\rlyapz \!\equiv\! \rlyap(\nu\!\!=\!\!0)$ at $\nu \!=\! 0$ 
gives us a measure of the conducting properties of the system.

In our bipartite hopping problem at $E \!=\! 0$,
we observe that the two sublattices decouple, and it is 
advantageous to transfer on one sublattice only (say, 
the $A$-sublattice), recovering the Lyapunov spectrum of the 
whole system from appropriate symmetries.
Furthermore, for the brickwall lattice of Fig.~\ref{brickwall}(a), 
the sublattice transfer in the $\hbx$ direction involves
only one column of $A$-sites at a time, but now the individual 
sublattice Lyapunov spectrum is not symmetric around zero.
(In passing we note that in such a system with a free left boundary,
there will be edge states living on the $A$-sites of the boundary,
and the number of such edge states is precisely the number of 
negative Lyapunov exponents for the transfer in this case.)
We employ all these simplifications in our numerical calculations; 
however, for the final results, we always quote the $\rlyapz$ 
for the whole system including both sublattices.

For the brickwall lattice of Fig.~\ref{brickwall}(a), there are 
two inequivalent directions $\hbx$ and $\hby$.  
However, we have checked that the system is either 
delocalized in both directions or is localized in both directions.
We therefore expect the system to exhibit true 2D behavior in the 
delocalized phase as long as one is not too close to the 
localization transition.
Our numerical results are given in the corresponding 
Figs.~\ref{HNWscan}~and~\ref{HNStDm}.
To estimate $\rlyapz$, we typically use several Lyapunov exponents 
that are closest in absolute value to zero, and we quote only 
the direction-averaged 
$\overline\rlyapz \!\equiv\! \sqrt{\rlyapz^{xx} \rlyapz^{yy}}\,$.
(If the Lyapunov exponents are defined as growth exponents per unit 
length instead of per lattice unit, and if the corresponding density 
is also defined per transverse length, the lattice spacings $a_x$ 
and $a_y$ enter both $\rlyapz^{xx} \sim a_x/a_y$ and 
$\rlyapz^{yy} \sim a_y/a_x$, but not $\overline\rlyapz$.)

We find that the system is indeed delocalized for $\delta\!=\!0$.
The fact that it remains delocalized for some range of values 
$\delta \!>\! 0$ can also be seen from the data at $\delta \!=\! 0$ 
by considering the transfer on the $A$-sublattice in the $\hbx$ 
direction:  
In this case, multiplying the horizontal bonds by $e^\delta$ 
simply shifts the whole Lyapunov spectrum rigidly down by 
$\delta$, $\nu_i \to \nu_i - \delta$.  Thus, there remains
some finite Lyapunov spectrum density $\rlyapz$ for a range 
of $\delta$, until the top of the spectrum reaches $\nu \!=\! 0$, 
and only then does the system localize.  The critical $\delta_c$ 
can be estimated very accurately, since we need to know only 
the largest Lyapunov exponent in the $\delta \!=\! 0$ case; 
for the particular system studied, we find 
$\delta_c \!\approx\! 1.432$.

\vskip 0.1cm
What ``conducting'' property of the system does the finite 
$\rlyapz$ correspond to?  Chalker and Bernhard\cite{ChaBer} 
suggest that $\rlyapz$ gives, up to some numerical factor, 
the conductivity of the 2D system in units of $e^2/h\,$:
\begin{equation}
\conduct \sim \rlyapz.
\label{sigma}
\end{equation}
However, we have to ask ourselves which conductivity is actually 
being ``measured'' by~Eq.~(\ref{sigma}), particularly since 
we are dealing with systems that have a singular DOS at the Fermi 
level.  In this respect, note that the above transfer matrix 
approach can be equivalently formulated as a recursion procedure 
for calculating the Green's function $(E-\hat H)^{-1}$ between 
the first and the last slices of the long strip,\cite{MacKra} 
while the largest localization length from such transfer matrix 
analysis corresponds to the average modulus-squared of such 
a ``propagator'' between the two slices.  
This transfer matrix estimate resembles the definition of 
conductivity used in the sigma-model literature (see, e.g.,
Refs.~\onlinecite{LudMpafShaGri}~and~\onlinecite{MckSto}), 
which is the conductivity we need in Eq.~(\ref{Ecross}).  
To make our estimates more consistent, we fix the proportionality 
factor in Eq.~(\ref{sigma}) by using the exact results for some
pure-systems.  Thus, an elementary calculation 
for the pure $\pi$-flux model gives $\rlyapz = 2/\pi$, 
which should be compared with $\conduct = 2 \times 1/\pi$ for 
the two Dirac points of the model's continuum limit 
(see Ref.~\onlinecite{LudMpafShaGri}).  From this and similar 
considerations for the brickwall lattice,\cite{brickLyap}
we conclude that the accurate relationship is simply 
$\conduct = \rlyapz$.

\vskip 0.1cm
Finally, a couple of asides.
The first concerns the random vector potential model,
for which Ref.~\onlinecite{LudMpafShaGri} predicts 
$\conduct = 2/\pi$ for {\em any} strength of the randomness.  
This result can be also obtained within the above transfer matrix
approach, if we assume that we can indeed fix the unknown 
numerical factor in Eq.~(\ref{Ecross}) to the appropriate constant 
value.  When we add the random surface $\Phi(\br)$ on top of, 
say, the pure $\pi$-flux model [see Eq.~(\ref{tPhi})],
the $E=0$ transfered wave functions need to be simply multiplied 
by the appropriate $e^{-\Phi(\alpha)}$ or $e^{\Phi(\beta)}$.
But this cannot change the Lyapunov spectrum of the transfer
along the quasi-$1$D strip for any finite transverse size of 
the strip, because $\Phi(\br)$ in such geometry is effectively 
a one-dimensional Gaussian surface and cannot fluctuate stronger 
than $\big[ \Phi(L_\parallel)-\Phi(0) \big]^2 \sim L_\parallel$ 
along the strip.  At this stage, it is not clear to us 
whether $\conduct$ should have any signature of the ``freezing'' 
transition at $g_c$ observed in the static and dynamic properties 
of this system; the above argument seems to suggest 
that there is none, but we cannot rule out the possibility that 
taking the limit $L_\parallel \to \infty$ while keeping $L_\perp$ 
fixed ``loses'' the information about the 2D system.

The second remark concerns the brickwall lattice of 
Fig.~\ref{brickwall}(a) and the transfer in the $\hbx$ direction.  
In this case, there is a direct correspondence between 
the conducting properties of the system as measured by the 
Lyapunov spectrum and the energetics of the domain walls 
in the corresponding dimer problem.  Such domain walls are defined 
relative to the fully locked state that obtains for 
$\delta>\delta_c$, and are forced to run in the $\hbx$ 
direction and cannot terminate; their number at a given $\delta$ 
is precisely the number of positive Lyapunov exponents 
in the hopping problem, while the free energy cost per 
unit length of adding or removing domain walls is precisely 
the spacing between the Lyapunov exponents near $\nu \!=\! 0$.

\subsection{Exact diagonalization studies: density of states}
\label{sec:NUMERICS:dos}
\begin{figure}
\epsfxsize=\columnwidth
\centerline{\epsffile{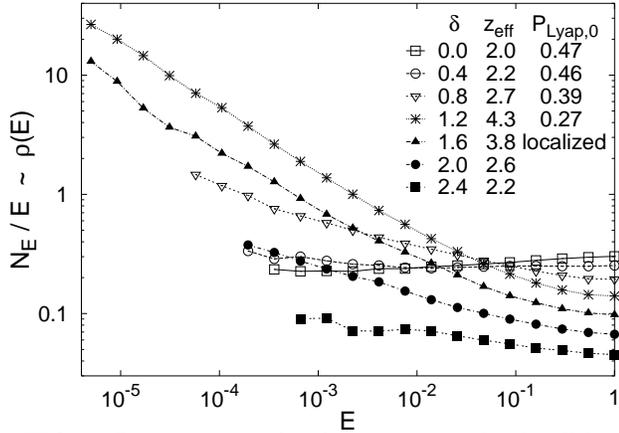}}
\caption{Density states for the anisotropic brickwall lattice
of Fig.~\ref{brickwall} calculated for $192 \times 192$ system 
with open bc in the $\hby$ direction and periodic bc in the $\hbx$ 
direction.  The regular (vertical) hopping amplitudes are chosen 
randomly from a uniform distribution in $[0,1]$, while the dimerized 
(horizontal) hopping amplitudes are chosen from $[0, e^\delta]$.  
The more detailed characterization of the data is similar to that 
of Fig.~\ref{HNWscan}.
}
\label{HNStDm}
\end{figure}

We also monitor the density of states as
we increase dimerization from $\delta \!=\! 0$, eventually driving 
the system into a localized state for $\delta \!>\! \delta_c$.
The corresponding results are shown in Fig.~\ref{HNStDm}, together 
with information about the system's localization properties.  
Consistent with our earlier discussion, the DOS divergence 
is seen to become 
stronger as we approach the localization transition, with the 
effective exponent $z_{\rm eff}(L\!\!=\!\!192)$ peaking at the 
transition.  Deep in the localized phase, the log-log plot of 
$\rho(E)$ vs $E$ is indeed a straight line, and the quoted values 
of $z$ give the actual bulk dynamical exponents.  
In the delocalized phase, on the other hand, there is still some 
curvature at the lowest energy scale, and the extracted values 
$z_{\rm eff}$ can only serve as rough indicators of the strength 
of the divergence.

\narrowtext

The actual data analysis is performed as follows:
In the delocalized phase, we fit the integrated density 
of states $N_E$ to the generalized-Gade form 
\begin{equation}
N_E = a \Gamma^{1/x}_E e^{-c \Gamma^{1/x}_E},
\label{fitGade}
\end{equation}
where $\Gamma_E \equiv \ln(\Omega_0/E)$.  Specifically, we try 
two values for the exponent $x$---the original Gade $x \!=\! 2$ and 
the modified $x \!=\! 3/2$.  The included prefactor $\Gamma^{1/x}_E$ 
is based on an assumption that there are no additional 
corrections to the asymptotic $E\to 0$ form 
$\rho(E) \sim E^{-1} \exp(-\Gamma_E^{1/x})$.  This is something that 
we do not really know---in fact, our lowest-gap analysis of 
Sec.~\ref{sec:BOUNDS:general} seems to suggest that there may be 
corrections stronger than any such $\Gamma^r$ prefactor.  
In this situation, we treat the above fit function only as a
baseline.  To be consistent, we should also allow for the 
uncertainty $\Gamma_E \to \Gamma_E + \Gamma_0$ 
in the bare energy scale relative to which the log-scale $\Gamma$ 
is defined.  As one might suspect, for a fixed fit range, we can 
approximate any DOS curve with the three parameters $a$, $c$, 
and $\Gamma_0$, so this sort of analysis is not really conclusive.
What we can do however is to look for the overall consistency of 
such fits:  
We can ask, for example, how the fit parameters change as we change 
the fit region, or how sensible the obtained parameters are 
(thus, the parameter $a$ should be of order one).  
Generally, the $x \!=\! 3/2$ form fares somewhat better in such 
analysis; also, the extracted numerical value of the parameter $c$ 
from such fits compares favorably with the lowest-gap results 
of the next subsection.  But this is as far as our direct DOS 
studies can take us in the delocalized phase.

In the localized phase, we have also tried fitting the DOS using 
a power-law times a logarithmic correction 
\mbox{$N_E \sim E^{d/z}\, (\Gamma_E)^r$}.  We typically
find that the best fit corresponds to $r \cong 0$, suggesting that 
the density of states has a simple power-law form in the localized 
phase.  This is also our conclusion from the lowest-gap studies 
presented below.

\subsection{The lowest energy state: exact diagonalization and 
dimer optimized defects studies}
\label{sec:NUMERICS:gap}
We can extend our numerical tests further by considering 
the distribution of the smallest positive energy in finite samples, 
paying particular attention to scaling with sample size.  
Such a direct diagonalization study can be performed for a factor of 
two larger systems than in our DOS studies, while indirect dimer 
methods can take us another factor of four in system size.  
This well-controlled and rather sensitive numerical approach
also corresponds closely to our analytical arguments in 
Sec.~\ref{sec:BOUNDS:general}.

We consider the same brickwall lattice system of 
Fig.~\ref{brickwall}(a); however, we now consider 
\mbox{$(2M_x+1) \!\times\! 2M_y$} ``odd$\,\times\,$even'' samples 
with spiral boundary conditions in the $\hbx$ direction 
and free boundaries in the $\hby$ direction.
The specific choice of the boundary conditions is such that
the dimer equivalence $\det\mt_{AB}=Z_d[A,B]$ holds precisely 
for this system;\cite{spiralbc} this allows us to compare the 
lowest-energy state of the hopping problem with the optimized 
defects in the corresponding dimer problem, exactly paralleling 
the discussion of Sec.~\ref{sec:BOUNDS:general}.  
To this end, we also calculate\cite{Mid00,optimize} the dimer RG 
``estimate'' Eq.~(\ref{Emin:dimerRG}) for the lowest gap, 
which is the quantity used in the bounds of 
Sec.~\ref{sec:BOUNDS:general}.

For the purposes of illustration, we show our results for two 
values of $\delta$, one in the delocalized phase ($\delta=0.8$) 
and one in the localized phase ($\delta=2.0$), both chosen 
well away from the transition point ($\delta_c = 1.432$).

\subsubsection{Delocalized phase, $\delta=0.8$}
\begin{figure}
\epsfxsize=\columnwidth
\centerline{\epsffile{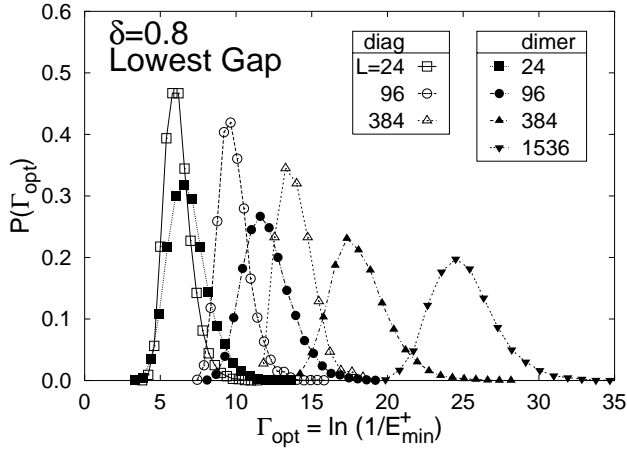}}
\vskip 0.2cm
\caption{``Lowest-gap'' analysis in the delocalized phase 
($\delta=0.8$) of the system of Fig.~\ref{brickwall}.
We consider $(L+1) \times L$ finite samples with spiral bc 
in the $\hbx$ direction and free bc in the $\hby$ direction.  
Open symbols show distributions of the smallest positive energy 
calculated by exact diagonalization methods, while filled symbols 
show distributions of the corresponding dimer RG ``estimate''
Eq.~(\ref{Emin:dimerRG}).
}
\label{gapdistHNStDm8}
\end{figure}

Fig.~\ref{gapdistHNStDm8} shows distributions of the logarithm
of the smallest positive energy in our samples in the delocalized 
phase.  The exact diagonalization results for 
$\Gopt = \ln (1/E_{\min}^{+})$ are shown with open 
symbols and extend up to system sizes $L=384$, while
the corresponding dimer RG estimates $ -\Eoptdef$ 
[cf.~Eq.~(\ref{Emin:dimerRG})]
are shown with filled symbols and extend up to $L=1536$.  
We see that the distributions move out strongly to larger absolute 
log-energies (i.e., smaller energies) with increasing $L$ and 
broaden somewhat at the same time.  We also see that the dimer RG 
estimates for the lowest gap are consistently smaller and
indeed provide strong lower bounds, at least for the larger systems. 
This is in agreement with our discussion in 
Sec.~\ref{sec:BOUNDS:general}.

\begin{figure}
\epsfxsize=\columnwidth
\centerline{\epsffile{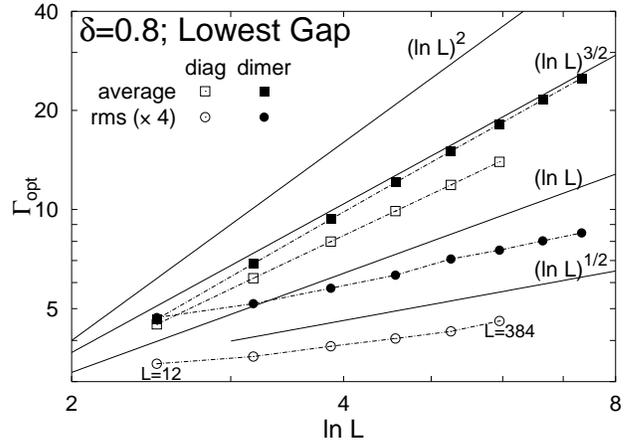}}
\vskip 0.2cm
\caption{Detailed analysis of the distributions of 
Fig.~\ref{gapdistHNStDm8}.  The mean $\overline\Gopt(L)$
and the standard deviation $\sigma(\Gopt;L)$ are plotted with 
open and filled symbols for the exact diagonalization and dimer RG 
results correspondingly.  Note the logarithmic scale for both 
the $\lnL$ and $\Gopt$ axes.  To gauge the $L$-dependence of 
$\overline\Gopt$ and $\sigma(\Gopt)$, we also plot different 
powers $(\ln L)^x$ with $x=2$, $3/2$, $1$, and $1/2$.
}
\label{gapanHNStDm8}
\end{figure}

In Fig.~\ref{gapanHNStDm8}, we analyze the behavior of the 
distributions $P(\Gopt;L)$ of Fig.~\ref{gapdistHNStDm8}, 
plotting their mean $\overline{\Gopt}(L)$ and the standard deviation 
$\sigma(\Gopt;L)$.  On a plot with linear scales for both 
$\lnL$ and $\Gopt$ (not shown), we observe a visible curvature
$\overline\Gopt \sim (\ln L)^x$ with $x>1$, indicative of a stronger 
than any finite-$z$ singularity.  In Fig.~\ref{gapanHNStDm8}, we use
logarithmic scales for both $\lnL$ and $\Gopt$ to get a rough 
estimate of the exponent $x$.  
The exact diagonalization results are seen to fall between 
the $x=1$ and $x=3/2$ lines, well away from the Gade $x=2$; 
at the same time, the optimized defect results, 
which extend to larger system sizes, clearly approach the $x=3/2$ 
line, as has been already checked in Ref.~\onlinecite{ZenLeaDsf} 
for a related model.
We also note a much weaker $\sim\! (\ln L)^{1/2}$ dependence of
the width of the distributions on the system size, consistent 
with the results of Ref.~\onlinecite{ZenLeaDsf} (summarized
in Sec.~\ref{sec:BOUNDS:general}).

In a more quantitative analysis, 
following Ref.~\onlinecite{ZenLeaDsf},
we perform linear fits for $[\overline\Gopt]^{2/3}$ and 
$[\sigma(\Gopt)]^2$ as functions of $\lnL$.  
For the exact diagonalization data, we obtain 
the asymptotic scaling
$\overline\Gopt \approx 0.81\, (\ln L)^{3/2}$ and
$\sigma \approx 0.42\, (\ln L)^{1/2}$.
We can compare this with the result from the DOS studies: 
performing the generalized-Gade fit Eq.~(\ref{fitGade}) of 
the data of Fig.~\ref{HNStDm}, we extract the parameter 
$c \cong 2.2$, which translates to a fairly close asymptotic 
scaling prediction $\overline\Gopt \approx 0.87\, (\ln L)^{3/2}$.  
Similarly, the dimer RG data yields 
$\overline\Gopt \approx 1.30\, (\ln L)^{3/2}$, not unreasonably
far from the above exact diagonalization predictions
[as discussed in Sec.~\ref{sec:BOUNDS:general}, we expect stronger 
subleading $O(\ln L)$ terms for the exact lowest gap $\Gopt$ than 
for the dimer $-\Eoptdef$].  Finally, we note that
Ref.~\onlinecite{ZenLea} calculated bulk defect density in 
a related vortex glass model as a function of the defect core energy 
(which corresponds to $\Gopt$ in the fermion problem), 
obtaining $N_\Gamma \sim \exp\left(-c \Gamma^{\,0.74}\right)$, 
fairly close to the expected 
$N_\Gamma \sim \exp\left(-c \Gamma^{\,1/x}\right)$ scaling form
with $x=3/2$.

\subsubsection{Localized phase, $\delta=2.0$}
\begin{figure}
\epsfxsize=\columnwidth
\centerline{\epsffile{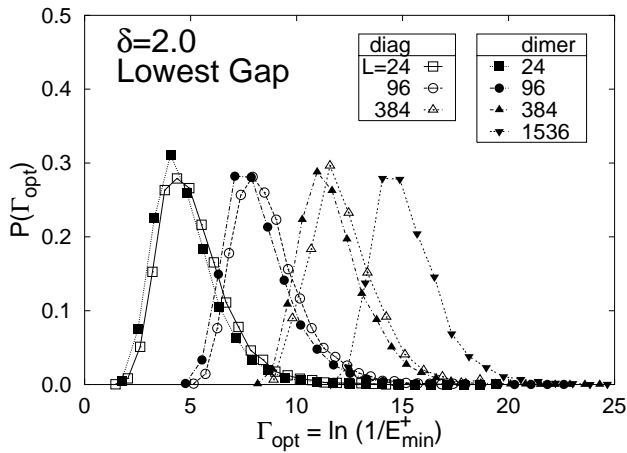}}
\vskip 0.2cm
\caption{Lowest-gap analysis in the localized phase, $\delta=2.0$,
similar to that of Fig.~\ref{gapdistHNStDm8}.
}
\label{gapdistHNStDm20}
\end{figure}

\begin{figure}
\epsfxsize=\columnwidth
\centerline{\epsffile{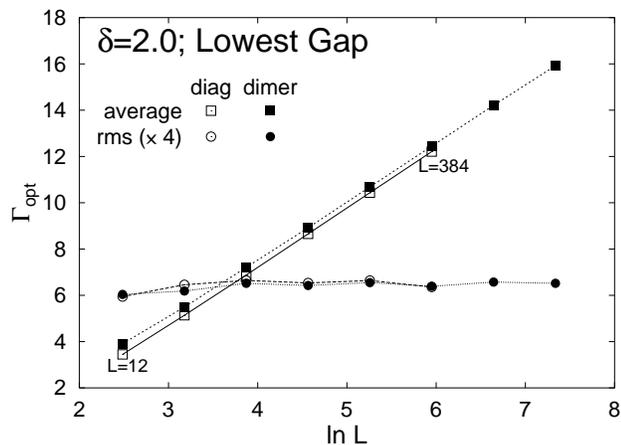}}
\vskip 0.2cm
\caption{Analysis of the distributions of Fig.~\ref{gapdistHNStDm20}.
}
\label{gapanHNStDm20}
\end{figure}

The lowest-gap analysis is simpler in the localized phase,
and is shown in Figs.~\ref{gapdistHNStDm20}-\ref{gapanHNStDm20}.
In this case too, the distributions move out to larger $\Gopt$
with increasing $L$, but now do so at a slower pace.  
We clearly observe simple finite $z$ scaling 
$\overline\Gopt = \const + z \lnL$, with $z\!=\!2.56$, which 
compares well with the estimate of $z\!=\!2.64$ from the DOS studies.
Also, the widths of the distributions essentially do not change 
with system size.  Note that the effects discussed in 
Sec.~\ref{sec:BOUNDS:general} produce a much smaller difference 
between the exact diagonalization results and the dimer RG estimates
in the localized phase, since the distance between the optimal 
defects is expected to scale very weakly $\sim\! \ln L$ with 
the sample size.  
Thus, at least at $\delta=2.0$ deep in the localized phase, and for 
the sizes studied, the dimer RG gives numerically accurate results 
for the lowest gap and the two peaks of the corresponding 
wave function in each sample.

\section{Discussion}
\label{sec:discus}
Our main results have already been summarized in the Introduction,
and here, we confine ourselves to some unresolved questions
in these bipartite models, as well as some issues in the closely
related ImRH problems.

Our first remark concerns the zero-energy wave function in the
``delocalized'' critical phase of the general bipartite problem.  
The multifractal properties of this wave function are expected 
to be trivial $\tau(q) \!\equiv\! 0$---this is because
of the flow to strong disorder $g\!\to\! \infty$
(see, e.g., Ref.~\onlinecite{CasChaFraGolMud}).
Equivalently, since the logarithm of the wave function magnitude 
is essentially the potential $\upsilon(\br)$ seen by the defects of 
the corresponding dimer problem [see Eq.~(\ref{vcor})], none of 
the participation ratios depends on system size---this is a 
consequence of the ground state dominance for this random potential.
In this respect, the critical wave function looks fairly localized.
The typical correlations like
$C_{\rm typ}(r) = 
\exp\Big[ \overline{\log|\Psi(\br) \Psi(\mb 0)|} \Big]$ are 
thus directly related to the correlations of $\upsilon(\br)$.  
On the other hand, the situation is less clear when we consider 
average correlations like
$C_{\rm av}(r) = \overline{|\Psi(\br) \Psi(\mb 0)|}$. 
Estimating these requires a better understanding of the statistics 
of such random surfaces.  For instance, it is not clear if 
$C_{\rm av}(r)$ is dominated by the near-returns of the surface 
to its global extrema, as is the case in the one-dimensional chain. 
Moreover, if this is indeed the case, what is 
the probability of finding such quasi-degenerate minima
at a fixed distance $r$?

Another feature of the delocalized phase that we have not discussed
is the suggested\cite{HwaDsf,ZenLeaHwa} universal susceptibility 
variations in the vortex glass problem, and it would be interesting 
to study the corresponding response in the fermionic problem in more 
detail.

The character of the transition between the delocalized and 
localized phases is a completely separate issue that we have 
ignored altogether.  It is not obvious if this critical end-point
of the line of fixed points that characterize the metallic phase 
exhibits dynamical scaling distinct from the metallic phase.
Another interesting feature of this transition
is that the conductivity $\conduct_{xx}$ remains finite (because
the Lyapunov spectrum density is finite at the top of the 
Lyapunov band), while $\conduct_{yy}$ vanishes.  It is not clear 
how this extreme anisotropy affects the properties of the transition,
and whether anything interesting remains.

In the localized phase, we have not discussed in detail
the structure of our low-energy ``string'' excitations.  
Presumably, they look similar to the domain walls (flux lines)
in the corresponding random dimer (vortex glass) problem.
Note that in the particular localized phase that we studied,
the strings are forced to run in the $\hbx$ direction.
In a more general localized phase, obtained by introducing
some other dimerization pattern, the situation is more 
complicated, but it still seems that it will be some kind of 
directed strings that will contribute most to the low-energy 
density of states:  The strings need to be stretched so that 
the end-to-end distance (setting the tunneling frequency scale) 
is of order the string length (determining the occurrence 
probability) for our counting arguments of Sec.~\ref{sec:RG:orig} 
to work.  An interesting system to study in this respect is a 
2D spinless superconductor with time-reversal invariance, 
which maps onto a bilayer BPRH.  
When the ``$p_x$-wave'' superconducting pairing amplitude 
$\Delta_{\mb j,\, \mb j + \mb x}$ is zero, 
the system is just a doubled Anderson localization problem, 
whereas increasing $\Delta$ from zero, we obtain a bipartite 
localized phase ``connected'' to the standard Anderson 
insulator point. 
Note the richness of the full phase diagram in this system:
a simple Lyapunov spectrum shift argument shows that increasing 
$\Delta$ further drives the system from this localized 
phase first into a critical delocalized phase, and then again into
a localized phase similar to our staggered band insulator!

\vskip 0.1cm
We conclude with some comments regarding more generic ImRH problems.
As is clear from the preceding discussion, we expect similar 
Griffiths string effects in the localized phases of such systems.
An important question is, of course, whether this Griffiths 
mechanism can compete with other mechanisms of ``filling the gap'' 
(which can, for example, produce a constant contribution to the 
density of states at the band center as in 
Ref.~\onlinecite{SenMpaf:noSR}).
In particular, are there situations with a power-law divergent 
density of states that has its origins in such Griffiths effects?
While we have not studied these questions in detail, there are 
certainly situations where this does indeed happen.  
Thus, in the particular fermionic ImRH
representation\cite{BlaPou} of the two-dimensional random bond 
Ising model, our preliminary results suggest that the density 
of states is power-law vanishing to the left (the less disordered
side) of the Nishimori line and power-law diverging to the right 
(this particular example was suggested to us by Nick Read and 
the corresponding result proved for a one-dimensional toy-model
in Ref.~\onlinecite{GruReaLud}).
Finally, 2D ImRH systems are believed to also have true metallic
phases,\cite{SenMpaf:noSR} and it would be interesting to consider 
these from this perspective.  In particular, can
one use an analogous dimer connection to gain further insight
into their properties?

\acknowledgments
We would like to thank N. Read for a useful discussion.
The work at Princeton was supported by NSF grant DMR-9802468,
and at Harvard by NSF grant DMR-9809483.

\appendix
\section{2D Gaussian surface}
\label{app:2dGAUSS}
The following facts are used in the main text to obtain bounds 
for the dynamical scaling in the random vector potential model; 
these are transcribed directly from Ref.~\onlinecite{CarDou01} 
for the two-dimensional Gaussian surface $\Phi(\mb r)$, 
Eq.~(\ref{Phi}), in an $L \times L$ box
(stipulating $\int\! d\br \, \Phi(\br) = 0$).

Extremal properties of the surface $\Phi(\br)$ are sharply defined: 
e.g., the maximum scales as
\begin{equation}
\Phi_{\max} (L) = 2\sqrt{\frac{g}{2\pi}} \ln L - 
\frac{3}{2}\sqrt{\frac{g}{8\pi}} \ln\ln L + \delta\Phi_{\max},
\end{equation}
with the sample-to-sample dependence entering only through 
an $O(1) \times \sqrt{g}$ random variable $\delta\Phi_{\max}$. 

The following ``partition function''
\begin{equation}
\FZ = \sum_\br e^{-2\Phi(\br)}
\label{ZLgauss}
\end{equation}
has a sharply defined logarithm:
\begin{eqnarray*}
\ln \FZ_g(L) & = & 2(1+\frac{g}{g_c}) \ln L + \Delta_g ~, 
~~~~~~~~~~~~~~ g<g_c \equiv 2\pi ~; \\
\ln \FZ_g(L) & = & 4 \ln L -\frac{1}{2}\ln\ln L + \Delta_g ~,
~~~~~~~~~~ g=g_c ~;\\
\ln \FZ_g(L) & = & \sqrt{\frac{g}{g_c}} (4 \ln L -\frac{3}{2}\ln\ln L)
+ \Delta_g ~,   
~~ g>g_c~;
\end{eqnarray*}
where $\Delta_g$ is an $O(1)$ random number (different in each case).
Note that for strong disorder $g>g_c$, $\FZ (L)$ is dominated by
the global minimum:
\begin{equation}
\FZ_g(L) = e^{-2\Phi_{\min}(L)} 
\sum_\br e^{-2 [\Phi(\br)-\Phi_{\min}(L)]},
\label{ZsumMax}
\end{equation}
where the sum contributes no $L$-dependence because 
$e^{-2 [\Phi(\mb r)-\Phi_{\min}]}$ remains normalizable
in the limit $L\to \infty$.


\begin{thebibliography}{10}
\bibitem{GraphZeroes}
If a given lattice does not allow complete matchings, as can happen, e.g., when
  the numbers of $A$ and $B$ sites are not equal, or when there are some
  disconnected or ``dangling'' lattice parts, the system will have exact
  zero-energy states, and the number of such ``graph zeros'' is equal to the
  minimal number of free sites that are left uncovered by any partial dimer
  cover of the system (also called ``deficiency'' number of the bipartite
  graph).
\bibitem{boundrg}
The leftmost inequality in Eq.~(\ref{Cnorm}) implies that the zero-energy
  RG~Eq.~(\ref{tprime2}) is always an upper bound for $E^+_{\min}$.
\bibitem{pbc1:short}
The brickwall RH problem [Fig.~\ref{brickwall}(a)] with free bc in the $\hbx$
  direction has of order $L_y$ edge states with exponentially small (in $L_x$)
  energies living on the free $A$ and $B$ sites of the correspondingly left and
  right boundaries. To study the bulk DOS, one would then have to separate the
  $O(L)$ edge and $O(L^2)$ bulk states in the total low-energy DOS, e.g., by
  varying the system size $L$. However, the bulk DOS can also be obtained more
  easily by having no free boundaries in the $\hbx$ direction, and this is what
  we do throughout.
\bibitem{myEcross}
Formula~(\ref{Ecross}) is taken verbatim from Ref.~\onlinecite{Gade},
  correcting for a minor inconsistency between Eqs.~(6.6)~and~(6.12) there.
\bibitem{brickLyap}
Consider the staggered square lattice system defined as $t_{\,\mb j,\, \mb j +
  \hbx} \!=\! t_x + (-1)^{j_x+j_y} \Delta_x$ and $t_{\,\mb j,\, \mb j + \hby}
  \!=\! t_y$, which for $t_x \!<\! t_y$ has two Dirac points, with $v_x \!=\! 2
  a_x \Delta_x$ and $v_y \!=\! 2 a_y \sqrt{t_y^2-t_x^2}$ [the brickwall lattice
  of Fig.~\ref{brickwall}(a) is a particular example with $\Delta_x \!\equiv\!
  t_x$]. For the transfer in the two directions we obtain $\rlyapz^{xx} \!=\!
  (v_x / v_y) \times 2/\pi$ and $\rlyapz^{yy} \!=\! (v_y / v_x) \times 2/\pi$,
  consistent with the anisotropic rescaling of the conductivity for the Dirac
  fermions in 2D.
\bibitem{spiralbc}
$Z_d[A,B]=\det \mt_{AB}$ for ``odd$\,\times\,$even'' brickwall lattice with
  spiral bc in the $\hbx$ direction and free bc in the $\hby$ direction because
  any transition cycle relative to the natural (horizontal) reference dimer
  cover winds precisely once along the $\hbx$ direction and therefore has odd
  length (Refs.\onlinecite{Kasteleyn,Montroll}).
\bibitem{optimize}
To find the optimized defects, we use a variant of the ``maximum-weight
  bipartite matching---shortest path'' combination of Ref.~\onlinecite{Mid00}.
\end{thebibliography}

\end{document}